\documentclass[a4paper,11pt]{article}

\pdfoutput=1 % if your are submitting a pdflatex (i.e. if you have
\usepackage{jheppub} % for details on the use of the package, please
\usepackage{slashed}
\usepackage{subcaption}
\usepackage{xcolor}

\usepackage[T1]{fontenc} % if needed

\newcommand{\cA}{\mathcal{A}}
\newcommand{\cM}{\mathcal{M}}
\newcommand{\cE}{\mathcal{E}}

\title{\boldmath Heavy Black Hole Effective Theory}

\author[]{Poul H. Damgaard,}
\author[]{Kays Haddad,}
\author[]{Andreas Helset}

\affiliation[]{Niels Bohr International Academy and Discovery Center,
Niels Bohr Institute, University of Copenhagen, Blegdamsvej 17,
DK-2100 Copenhagen, Denmark}

\emailAdd{phdamg@nbi.ku.dk}
\emailAdd{kays.haddad@nbi.ku.dk}
\emailAdd{ahelset@nbi.ku.dk}

\abstract{We formulate an effective field theory describing large mass scalars and fermions minimally coupled to gravity. The operators of this effective field theory are organized in powers of the transfer momentum divided by the mass of the matter field, an expansion which lends itself to the efficient extraction of classical contributions from loop amplitudes in both the post-Newtonian and post-Minkowskian regimes. We use this effective field theory to calculate the classical and leading quantum gravitational scattering amplitude of two heavy spin-1/2 particles at the second post-Minkowskian order.}
\preprint{SAGEX-19-20-E}

\allowdisplaybreaks

\begin{document} 
\maketitle
\flushbottom

\section{Introduction}
\label{sec:intro}

The direct detection of gravitational waves (GWs) from the merging of two black holes by LIGO and VIRGO in 2015 \cite{LIGOGW} has placed a spotlight on GW astronomy as a novel channel through which to test general relativity (GR). As the detection rate of GWs becomes more frequent in the years ahead, it is necessary to improve the analytical predictions on which the GW templates used in the observations are based. To do so requires knowledge of the interaction Hamiltonian of a gravitationally bound binary system to high accuracy. This necessarily entails the calculation of higher orders in the post-Newtonian (PN) and post-Minkowskian (PM) expansions.

Much of the work related to GWs has been done from the relativistic approach to GR; some notable developments are the effective-one-body approach  \cite{Buonanno:1998gg,Damour:2016gwp,Damour:2017zjx}, numerical relativity \cite{Baker:2005vv,Campanelli:2005dd,Pretorius:2005gq}, and effective field theoretic methods \cite{Goldberger:2004jt,Porto:2005ac} %, which are useful in describing the inspiral and merger phases respectively of a binary black hole system 
(see Refs.~\cite{Porto:2016pyg,Levi:2018nxp} for comprehensive reviews summarizing most of the analytical aspects of these methods). Also, there has been substantial work done using traditional and modern scattering amplitude techniques to calculate classical gravitational quantities, including the non-relativistic classical gravitational potential \cite{Donoghue:1993eb,Donoghue:1994dn,BjerrumBohr:2002kt,Holstein:2008sx,Bjerrum-Bohr:2013bxa,Cachazo:2017jef,Guevara:2017csg,GRSA,Guevara:2018wpp,Chung:2018kqs,Zvi3PM,Guevara:2019fsj,Bern:2019crd}. Moreover, techniques were recently presented in Refs.~\cite{Cheung:2018wkq,Cristofoli:2019neg} to convert fully relativistic amplitudes for scalar-scalar scattering to the classical potential, and for obtaining the scattering angle directly from the scattering amplitude \cite{Kalin:2019rwq}.
The prescription of Ref.~\cite{Cheung:2018wkq} was combined there with modern methods in amplitude computations to obtain the 2PM, and elsewhere the state-of-the-art 3PM Hamiltonian for classical scalar-scalar gravitational scattering \cite{Zvi3PM,Bern:2019crd}.
This large body of work, facilitated by classical effects arising at all loop orders \cite{Donoghue:1993eb,Holstein:2004dn} (see Section \ref{sec:hbarCount}), suggests that quantum field theory methods can reliably be used instead of direct computation from GR, particularly when the latter becomes intractable. Following in this vein, we apply here the machinery of effective field theory (EFT) to compute classical gravitational scattering amplitudes.

Computations of classical quantities from quantum scattering amplitudes are inherently inefficient. Entire amplitudes must first be calculated --- which are comprised almost entirely of quantum contributions --- and then classical terms must be isolated in a classical limit. One of the advantages of EFT methods is that they allow the contributions of certain effects to be targeted in amplitude calculations, thus excluding terms that are not of interest from the outset. From the point of view of classical gravity, it is then natural to ask whether an EFT can be formulated that isolates classical from quantum contributions already at the operator level. Indeed, we find that a reinterpretation of the operator expansion of the well-established Heavy Quark Effective Theory (HQET) \cite{Georgi:1990um,Bodwin:1994jh} (for a review, see, {\it e.g.}, Ref.~\cite{HQETRev}) leads us down the right path.

HQET has been used extensively to describe bound systems of one heavy quark --- with mass $M$ large relative to the QCD scale $\Lambda_{\text{QCD}}$ --- and one light quark --- with mass $m\lessapprox\Lambda_{\text{QCD}}$. Interactions between the light and heavy quarks are on the order of the QCD scale, $q\sim\Lambda_{\text{QCD}}$. Thus the heavy quark can, to leading order, be treated as a point source of gluons, with corrections to the motion of the heavy quark arising from higher-dimensional effective operators organized in powers of $q/M\sim\Lambda_{\text{QCD}}/M$.

A similar hierarchy of scales exists when considering the long-range (classical) gravitational scattering of two heavy bodies; for long-range scattering of macroscopic objects the momentum of an exchanged graviton $q$ is much smaller than the mass of each object. This can be seen by noting that, \iffalse since the exchange of a boson is a quantum process,\fi once powers of $\hbar$ are restored, the transfer momentum is $q=\hbar\bar{q}$, where $\bar{q}$ is the wavenumber of the mediating boson \cite{Kosower:2018adc}. Consequently, the expansion parameter of HQET --- and its gravitational analog, which we refer to as the Heavy Black Hole Effective Theory (HBET) --- can be recast as $\hbar\bar{q}/M$. The magnitude of the wavenumber is proportional to the inverse of the separation of the scattering bodies, hence for macroscopic separations and masses, $\hbar\bar{q}/M\ll1$. The presence of this separation of scales in classical gravitational scattering further motivates the development of HBET. The explicit $\hbar$ power counting of its operators makes HBET a natural framework for the computation of classical gravitational scattering amplitudes.

This work shares conceptual similarity with the Non-Relativistic General Relativity (NRGR) EFT approach to the two-body problem introduced in Ref.~\cite{Goldberger:2004jt} (extended to the case of spinning objects in Ref.~\cite{Porto:2005ac}). As in the case of NRGR, the interacting objects of HQET and HBET are sources for the mediating bosons, and are not themselves dynamical; in HQET and HBET, this can be seen from the fact that derivatives in the Lagrangians produce residual momenta (see Sec.~\ref{sec:HQET}) in the Feynman rules, not the full momenta of the objects in the scattering. However the EFTs differ in what they describe. NRGR is organized in powers of velocity, facilitating the computation of the Post-Newtonian expansion. In contrast, the operator expansions of HQET and HBET are expansions in $\hbar$, allowing us to target terms in the amplitudes with a desired $\hbar$ scaling. Being derived directly from a relativistic quantum field theory, a Post-Minkowskian expansion is naturally produced by the amplitudes of HBET. Moreover, while NRGR computes the non-relativistic interaction potential directly, HBET is intended for the computation of the classical portions of scattering amplitudes, which must then be converted to classical observables \cite{Cheung:2018wkq,Cristofoli:2019neg,Kalin:2019rwq}.

In this paper, we derive HBET in two forms, describing separately the interactions of large mass scalars and fermions minimally coupled to gravity. By restoring $\hbar$ we demonstrate how to determine which operators contribute classically to $2\rightarrow2$ scattering at $n$ loops. Using the developed EFT we compute the $2\rightarrow2$ classical scattering amplitude for both scalars and fermions up to 2PM order. We include in our calculations the leading quantum contributions to the amplitudes that originate from the non-analytic structure of the loop integrals.

The structure of this paper is as follows. In Section~\ref{sec:hbarCount} we explain the procedure by which we restore $\hbar$ in the amplitudes. We give a brief review of HQET in Section~\ref{sec:HQET}, and outline the derivation of the HQET Lagrangian. Our main results are presented in Sections~\ref{sec:HBET} and \ref{sec:Amps}. In the former we derive the HBET Lagrangians for heavy scalars and heavy fermions, whereas the latter presents the $2\rightarrow2$ scattering amplitudes for each theory up to 2PM. We conclude in Section~\ref{sec:Conclusions}. Technical details of the 
HQET spinors are discussed in Appendix~\ref{sec:HeavySpinors}. In Appendix~\ref{sec:HSQED} we include the effective theory of a heavy scalar coupled to electromagnetism, and in Appendix~\ref{sec:AmpsEM} we use HQET to compute the classical and leading quantum contributions to the $2\rightarrow2$ electromagnetic amplitude up to one-loop. Appendices~\ref{sec:FeynRules} and \ref{sec:Integrals} contain respectively the Feynman rules and a discussion on the one-loop integrals needed to perform the 2PM calculations. We also discuss in Appendix~\ref{sec:Integrals} the circumvention of the so-called pinch singularity, which appears in some HQET loop integrals.%\footnote{Loop amplitudes in HQET are known to suffer from the so-called pinch singularity when describing a bound state of two heavy quarks. We circumvent this issue by treating only the scattering of two unbound heavy objects. This point is discussed further in Appendix \ref{sec:Integrals}.} 

%%%%%%%%%%%%%%%%%%%%%%%%%%%%%%%%%
%   Counting  hbar
%%%%%%%%%%%%%%%%%%%%%%%%%%%%%%%%%

\section{Counting \texorpdfstring{$\hbar$}{h-bar}}
\label{sec:hbarCount}

In quantum field theory we are accustomed to working with units where both the reduced Planck 
constant $\hbar$ and the speed of light $c$ are set to unity, thus obscuring the classical limit $\hbar\rightarrow0$.
We must therefore systematically restore the powers of $\hbar$ in scattering amplitudes so that a classical limit may be taken. We follow Ref.~\cite{Kosower:2018adc} to do so.

The first place we must restore $\hbar$ is in the coupling constants such that their dimensions remain unchanged: in both gravity and QED, the coupling constants are accompanied by a factor of $\hbar^{-1/2}$. 
Second, as mentioned above, we must distinguish between the momentum of a massless particle $p^{\mu}$ and its wavenumber $\bar{p}^{\mu}$. They are related through
\begin{align}
    \label{eq:MomWaveVec}
    p^{\mu}&=\hbar\bar{p}^{\mu}.
\end{align}
In the classical limit, the momenta and masses of the massive particles must be kept constant, whereas for massless particles it is the wavenumber that must be kept constant. While this result is achieved formally through the consideration of wavefunctions in Ref.~\cite{Kosower:2018adc}, an intuitive way to see this is that massless particles are classically treated as waves whose propagation can be described by a wavenumber, whereas massive particles are treated as point particles whose motion is described by their momenta.

In this work, we are interested in the scattering of two massive particles, where the momentum $q$ is transferred via massless particles (photons or gravitons). Letting the incoming momenta be $p_1$ and $p_2$, the amplitudes will thus take the form
\begin{equation}
	\label{eq:ampkin}
	i\mathcal{M}(p_1, p_2 \rightarrow p_1 - \hbar \bar q, p_2 + \hbar \bar q).
\end{equation}
As the momentum transfer is carried by massless particles, the wavenumber $\bar q$ remains fixed in the classical limit, whereas the momentum $q$ scales with $\hbar$, as indicated in Eq.~\eqref{eq:ampkin}. The classical limit of the kinematics is therefore associated with the limit $|q|\rightarrow0$.

\subsection{Counting at one-loop}

With these rules for restoring powers of $\hbar$ in amplitudes, we can preemptively deduce which operators from the EFT expansion can contribute classically at one-loop level. First we must determine the $\hbar$-scaling that produces classical results.

The usual Newtonian potential can be obtained from the Fourier transform of the leading order non-relativistic contribution to the tree-level graviton exchange amplitude (see Fig.~\ref{fig:tChan}). Using a non-relativistic normalization of the external states,
\begin{align}
    \langle p_{1}|p_{2}\rangle&=(2\pi)^{3}\delta^{3}(\vec{p}_{1}-\vec{p}_{2}),
\end{align}
this contribution to the amplitude is
\begin{align}
    \cM^{(1)}&\approx-\frac{\kappa^{2}m_{1}m_{2}}{8q^{2}},
\end{align}
where $\kappa=\sqrt{32\pi G/\hbar}$ and $G$ is Newton's constant. Here $q$ is the four-momentum of the mediating graviton. Following the discussion above, we can thus make all factors of $\hbar$ explicit by writing $q$ in terms of the graviton wavenumber. We find
\begin{align}
    \cM^{(1)}&\approx-\frac{4\pi Gm_{1}m_{2}}{\hbar^{3}\bar{q}^{2}}\label{eq:ClassicalAmp}.
\end{align}
We conclude that classical contributions to scattering amplitudes in momentum space with the current conventions scale as $\hbar^{-3}$. A quantum mechanical term is thus one that scales with a more positive power of $\hbar$ than this, as such a term will be less significant in the $\hbar\rightarrow0$ limit.

Indeed, this must be the $\hbar$-scaling of any term in the amplitude contributing classically to the potential. At tree-level, the relation between the amplitude and the potential is simply
\begin{align}
    V&=-\int\frac{d^{3}q}{(2\pi)^{3}}e^{-\frac{i}{\hbar}\vec{q}\cdot\vec{r}}\cM =-\hbar^{3}\int\frac{d^{3}\bar{q}}{(2\pi)^{3}}e^{-i\vec{\bar{q}}\cdot\vec{r}}\cM,
\end{align}
where we have made factors of $\hbar$ explicit. The scaling of classical contributions from the amplitude must be such that they cancel the overall $\hbar^{3}$ in the Fourier transform.

Central to the applicability of the Feynman diagram expansion to the computation of classical corrections to the interaction potential is the counterintuitive fact that loop diagrams can contribute classically to scattering amplitudes \cite{Donoghue:1993eb,Holstein:2004dn}. Which loop diagrams may give rise to classical terms can be determined by requiring the same $\hbar$-scaling as in Eq.~\eqref{eq:ClassicalAmp}.

Diagrams at one-loop level have four powers of the coupling constant, which are accompanied by a factor of $\hbar^{-2}$. This implies that classical contributions from one-loop need to carry exactly one more inverse power of $\hbar$, arising from the loop integral. The only kinematic parameter in the scattering that can bring the needed $\hbar$ is the transfer momentum $q$, and even then only in the non-analytic form $1/\sqrt{-q^{2}}$. Non-analytic terms at one-loop arise from one-loop integrals with two massless propagators \cite{Donoghue:1993eb,Holstein:2004dn}. There are three topologies at one-loop that have two massless propagators per loop, and hence three topologies from which the requisite non-analytic form can arise: the bubble, triangle, and (crossed-)box topologies. We will determine the superficial $\hbar$-scaling of these topologies.

First we note that the loop momentum $l$ can always be assigned to a massless propagator, and hence should scale with $\hbar$. The bubble integral is thus
\begin{align}
	i\mathcal{M}_{\rm bubble}^{(2)} &\sim \frac{G^2}{\hbar^2} \hbar^4 \int d^4\overline{l}
	\frac{1}{\hbar^2 \overline{l}^2} \frac{1}{\hbar^2 (\overline{l}+\overline{q})^2} +
	\mathcal{O}(\hbar^{-1})\notag \\
	&=\mathcal{O}(\hbar^{-2}).
\end{align}
We conclude that the bubble contains no classical pieces.

Triangle integrals must have an extra HQET/HBET matter propagator, which, as will be seen below, is linear in the residual momentum. Therefore, triangle diagrams scale as
\begin{align}
	i\mathcal{M}_{\rm triangle}^{(2)} &\sim \frac{G^{2}}{\hbar^2} \hbar^4 \int d^4\overline{l}
	\frac{1}{\hbar^2 \overline{l}^2} \frac{1}{\hbar^2 (\overline{l}+\overline{q})^2}\frac{1}{\hbar v\cdot (\overline{l} + \overline{k})} +
	\mathcal{O}(\hbar^{-2})\notag \\
	&=\mathcal{O}(\hbar^{-3}).
\end{align}
Here, $v$ is the velocity of the heavy quark, and $k$ is the residual HQET/HBET momentum. These quantities and their $\hbar$-scaling are discussed in Section \ref{sec:HQET}. The scaling of the triangle integral suggests that triangles must contain classical pieces.

Finally, box and crossed-box integrals scale as
\begin{align}
	i\mathcal{M}_{\rm (crossed-)box}^{(2)} &\sim \frac{G^{2}}{\hbar^2} \hbar^4 \int d^4\overline{l}
	\frac{1}{\hbar^2 \overline{l}^2} \frac{1}{\hbar^2 (\overline{l}+\overline{q})^2}
	\frac{1}{\hbar v\cdot (\overline{l} + \overline{k}_{1})}
	\frac{1}{\hbar v\cdot (\overline{l} + \overline{k}_{2})} +
	\mathcal{O}(\hbar^{-3})\notag \\
	&=\mathcal{O}(\hbar^{-4}).
\end{align}
There are potentially classical pieces in the subleading terms of the (crossed-)box -- that is, in higher rank (crossed-)box loop integrals. However, the leading terms in the box and crossed-box diagrams look to be too classical, scaling as $1/\hbar^4$.
In order for the amplitude to have a sensible classical limit, such contributions must cancel in physical classical quantities. Two types of cancellations occur at one-loop level: cancellations between the box and crossed-box, and cancellations due to the Born iteration of lower order terms when calculating the potential \cite{Holstein:2008sw,Holstein:2008sx,Neill:2013wsa,Cristofoli:2019neg}.

In this paper we compute long-range effects
%the classical contributions 
arising from one-loop integrals, which are proportional to the non-analytic factors $S\equiv\pi^{2}/\sqrt{-\bar{q}^{2}}$ %, and the leading quantum contributions, proportional to the non-analytic factor 
and $L\equiv\log(-q^{2})$.\footnote{In contrast to Refs.~\cite{Holstein:2008sx,Holstein:2008sw}, we define $S$ in terms of the wavenumber $\bar{q}$ to make powers of $\hbar$ explicit in the amplitude.} When considering only spinless terms at one-loop order, those proportional to $S$ are classical, and those proportional to $L$ are quantum. With the established $\hbar$ counting, classical terms at one-loop can arise from operators with at most one positive power of $\hbar$, and quantum terms arise from operators with at most two positive powers of $\hbar$. In the operator expansion of HQET/HBET, powers of $\hbar$ come from partial derivatives.

The inclusion of spin slightly complicates this counting. In order to identify spin multipoles with those of the classical angular momentum, we must allow the spin to be arbitrarily large while simultaneously taking the classical limit. More precisely, for a spin $S_{i}$ the simultaneous limits $S_{i}\rightarrow\infty$, $\hbar\rightarrow0$ must be taken while keeping $\hbar S_{i}$ constant \cite{Maybee:2019jus,Chung:2019duq}.\footnote{The universality of the multipole expansion in gravitational interactions ensures that the expansion remains unchanged in this limit \cite{Holstein:2008sx,Chung:2019duq}.} When considering spin-inclusive parts of the amplitude we must therefore neglect one positive power of $\hbar$ for each power of spin when identifying the classical and quantum contributions. To make the expansion in classical operators explicit, in this paper we keep track only of the factors of $\hbar$ that count towards the determination of the classicality of terms in the amplitudes. Practically, this amounts to rescaling the Dirac sigma matrices in the operators as $\sigma^{\mu\nu}\rightarrow \sigma^{\mu\nu}/\hbar$, or the spins in the amplitudes as $S_{i}\rightarrow S_{i}/\hbar$. At linear order in spin, this leads again to the interpretation at 2PM order of terms proportional to $S$ as being classical, and those proportional to $L$ as being quantum. At quadratic order in spin, however, terms such as $q^{3}S$ and $qL$ begin arising, which respectively have quantum and classical $\hbar$ scaling. 

Altogether, operators contributing classically contain either up to one deriative, or up to two derivatives and a Dirac sigma matrix, which will be seen to be related to the spin vector.

\subsection{Counting at \texorpdfstring{$n$}{n}-loops}

We can extend this analysis to determine which operators can produce classical terms at arbitrary loop order. First we consider two-loop diagrams, contributing 3PM corrections to the classical potential.

The highest order operator needed is determined by the most classical $\hbar$-scaling attainable at a given loop order, \textit{i.e}, by the $\hbar$-scaling of the diagram that scales with the most inverse powers of $\hbar$. In Appendix \ref{sec:FeynRules} we show that the leading order $\hbar$-scaling of a graviton-matter vertex is always $\hbar^{0}$. The Einstein-Hilbert action governing pure graviton vertices involves two derivatives of the graviton field, so that pure graviton vertices always scale as $\hbar^{2}$. It follows that the most classical diagrams at 3PM are the box and crossed-box with four massive and three massless propagators; we refer to these as ladder diagrams. The overall coupling is $G^3/\hbar^3$, and the two integrals over loop momenta contribute eight positive powers of $\hbar$. In total, the amplitude superficially scales as
\begin{align}
	\cM^{(3)}_{\textrm{ladder}} \sim \frac{1}{\hbar^3}\hbar^8\frac{1}{\hbar^{10}} = \frac{1}{\hbar^5}.
\end{align}
%
\iffalse
That is too superclassical. However, we did not take into account the numerators. For interactions
between massive and massless particles, the interaction is of order 1 ($v^\mu$). For the 
interactions between purely massless particles, the interaction is of order $\hbar$, ($q^\mu$).
The double box with two massive propagators have two interactions with purely massless 
particles. Thus, the overall scaling is 
%
\begin{align}
	\cA_{\textrm{3PM Box}} \sim \frac{1}{\hbar^5}.
\end{align}
%
This is the same scaling as the double box with four massive propagators. In that case, 
there are no interactions among
the massless particles, and two fewer massless propagators, and two more massive propagators.
\fi

At $n$PM --- corresponding to $n-1$ loops --- the dominant diagrams in the $\hbar\rightarrow0$ limit are still the ladder diagrams, with $n$ massless propagators and $2(n-1)$ massive propagators. The scaling is then
\begin{align}
	\cM^{(n)}_{\textrm{ladder}} &\sim \frac{1}{\hbar^{n}}\hbar^{4(n-1)}\frac{1}{\hbar^{2n}}\frac{1}{\hbar^{2(n-1)}}\notag \\
	&\sim\frac{1}{\hbar^{2+n}}.
\end{align}
%
%The counting of the other diagrams follow straightforwardly.
From the HBET point of view, this means that we need to include operators that
scale with one more power of $\hbar$ whenever we go from $n$PM to $(n+1)$PM order. Starting with the observation from the previous section of classical operators at 2PM, we will need operators with at most $n-1$ derivatives, or $n$ derivatives and one Dirac sigma matrix, to obtain the full classical correction at $n$PM. Furthermore, to have a sensible classical limit, all superclassical contributions must cancel in physical quantities. We see that the order of cancellation scales with the number of loops.

Note that, according to this counting, starting at 3PM, spinless terms proportional to $L$ can contribute classically. This is consistent with the classical 3PM scalar-scalar amplitude in Refs.~\cite{Zvi3PM,Bern:2019crd}.

\section{Heavy Quark Effective Theory}
\label{sec:HQET}

As the concepts and methods we will use to derive HBET are based on those of HQET, we give a brief review of the latter here.

HQET is used in calculations involving a bound state of a heavy quark $Q$ with mass $m_{Q}\gg \Lambda_{\text{QCD}}$, 
and a light quark with mass smaller than $\Lambda_{\text{QCD}}$. 
The energy scale of the interactions between the light and heavy quark is on the order of the QCD scale, 
and is thus small compared to the mass of the heavy quark. 
The momentum $p^{\mu}$ of the system is decomposed into a large part, $m_{Q}v^{\mu}$, and a small residual momentum, $k^{\mu}$. Altogether,
\begin{align}
	p^{\mu}=m_{Q}v^{\mu}+k^{\mu},\ \qquad |k^{\mu}|\sim\mathcal{O}(\Lambda_{\text{QCD}}) \,\,\, 
\text{where} \,\,\, \Lambda_{\text{QCD}} \ll m_Q. \label{eqn:HQETMomDecomposition}
\end{align}
A hierarchy of scales is present, and we can organize an effective theory which expands in this
hierarchy.

An interesting feature of HQET, as will be seen below, is that its propagating degrees of freedom are massless. The propagating degrees of freedom carry the residual momentum $k^{\mu}$. Therefore, since we are interested in classical scattering, we can rewrite the residual momentum according to Eq.~\eqref{eq:MomWaveVec}:
\begin{align}
	p^\mu = m_Q v^\mu + \hbar \bar{k}^\mu.
\end{align}

The procedure we will use to derive the HBET Lagrangian for spinors in the next section is identical to that used to derive the HQET Lagrangian. As such, we outline the derivation of the HQET Lagrangian for one quark coupled to a $U(1)$ gauge field.\footnote{The non-abelian case is discussed in {\it e.g.} Ref.~\cite{ManoharHQET}.} Our starting point is the QED Lagrangian,
\begin{align}
	\mathcal{L}_{\text{QED}}&=\bar{\psi}\left(i\slashed{D}-m\right)\psi,\ \qquad \textrm{where} \quad D^{\mu}\psi\equiv(\partial^{\mu}+ieA^{\mu})\psi. \label{eqn:QEDLag}
\end{align}
Next, following the pedagogical derivation in Ref.~\cite{Schwartz:2013pla}, we introduce the projection operators
\begin{subequations}
\begin{align}
    P_{\pm}\equiv\frac{1\pm\slashed{v}}{2},
\end{align}
and two eigenfunctions of these operators
\begin{align}
    Q&\equiv e^{imv\cdot x}P_{+}\psi,\label{eq:HQETSpinor} \\
    \tilde{Q}&\equiv e^{imv\cdot x}P_{-}\psi\label{eq:HQETAntiSpinor}.
\end{align}
\end{subequations}
This allows us to decompose the spinor field as
\begin{align}
\psi&=\frac{1+\slashed{v}}{2}\psi+\frac{1-\slashed{v}}{2}\psi 
= e^{-imv\cdot x}\left(Q+\tilde{Q}\right). \label{eqn:HQETSpinor}
\end{align}
The details pertaining to the external states of the fields $Q$ and $\tilde{Q}$ are discussed in Appendix~\ref{sec:HeavySpinors}.

Substituting Eq.~\eqref{eqn:HQETSpinor} into Eq.~\eqref{eqn:QEDLag}, using some simple gamma matrix and projection operator identities,
and integrating out $\tilde{Q}$ using its equation of motion, we arrive at the HQET Lagrangian,
\begin{align}
	\mathcal{L}_{\text{HQET}}&=\bar{Q}\left(iv\cdot D-\frac{D^{2}_{\perp}}{2m}-\frac{e}{4m}\sigma^{\mu\nu}F_{\mu\nu}\right)Q
	+\frac{1}{2m}\bar{Q}i\slashed{D}_{\perp}\sum^{\infty}_{n=1}\left(-\frac{iv\cdot D}{2m}\right)^{n}i\slashed{D}_{\perp}Q. \label{eq:HQETLag}
\end{align}
Here, $\sigma^{\mu\nu}\equiv\frac{i}{2}[\gamma^{\mu},\gamma^{\nu}]$ is the Dirac sigma matrix and $D^{\mu}_{\perp} \equiv D^{\mu} - v^{\mu} (v\cdot D)$ is the
covariant derivative orthogonal to $v^{\mu}$.

The redundant operators proportional to the leading order equation of motion can be removed by the field redefinition \cite{ManoharHQET}
\begin{align}
	Q\rightarrow&\left(1-\frac{D^{2}_{\perp}}{8m^{2}}-\frac{e}{16m^{2}}\sigma_{\mu\nu}F^{\mu\nu}+\frac{1}{16m^{3}}D^{\mu}_{\perp}(iv\cdot D)D_{\perp\mu}-\frac{e}{16m^{3}}v_{\mu}D_{\perp\nu}F^{\mu\nu}\right. \notag\\
		    &\quad\left.-\frac{i}{16m^{3}}\sigma_{\mu\nu}D^{\mu}_{\perp}(iv\cdot D)D^{\nu}_{\perp}-\frac{ie}{16m^{3}}v_{\rho}\sigma_{\mu\nu}D^{\mu}_{\perp}F^{\nu\rho}\right)Q
\end{align}
to order $\mathcal{O}(1/m^4)$, leading to the Lagrangian
\begin{align} \label{eq:HQETLagRedef}
	\mathcal{L}_{\text{HQET}}&=\bar{Q}\left(iv\cdot D-\frac{D^{2}_{\perp}}{2m}+\frac{D^{4}_{\perp}}{8m^{3}}-\frac{e}{4m}\sigma_{\mu\nu}F^{\mu\nu}-\frac{e}{8m^{2}}v^{\mu}[D^{\nu}_{\perp}F_{\mu\nu}]\right.\\
				 &\quad\left.+\frac{ie}{8m^{2}}v_{\rho}\sigma_{\mu\nu}\{D^{\mu}_{\perp},F^{\rho\nu}\}+\frac{e}{16m^{3}}\{D^{2}_{\perp},\sigma_{\mu\nu}F^{\mu\nu}\}+\frac{e^{2}}{16m^{3}}F_{\mu\nu}F^{\mu\nu}\right)Q+\mathcal{O}(m^{-4}). \nonumber
\end{align}
Square brackets enclosing a derivative denote that the derivative acts only within the brackets.

Once Fourier transformed, partial derivatives produce the momentum of the differentiated field% divided by $\hbar$, cancelling the $\hbar$ from the Eq.~\eqref{eqn:QEDLag}
. In the specific case of HQET, the partial derivatives produce either a residual momentum (when acting on the spinor field) or a photon momentum (when acting on the vector field) in the Feynman rules. As both types of momenta correspond to massless modes, they both scale with $\hbar$, and hence partial derivatives always result in one positive power of $\hbar$.

\section{Heavy Black Hole Effective Theory}
\label{sec:HBET}
We now turn to the case of a heavy particle minimally coupled to gravity. The derivation of the Lagrangian for a heavy scalar coupled to gravity differs from the derivation of the spinor theory, because the scalar field whose heavy-mass limit we are interested in describing is real. The initial Lagrangian is that of a minimally coupled scalar matter field:
\begin{align}
	\sqrt{-g} \mathcal{L}_{\text{sc-grav}}&=\sqrt{-g}\left(\frac{1}{2}g^{\mu\nu}\partial_{\mu}\phi\partial_{\nu}\phi-\frac{1}{2}m^{2}\phi^{2}\right). \label{eqn:ScGravLag}
\end{align}
The metric is given by a small perturbation around flat space, $g_{\mu\nu}=\eta_{\mu\nu}+\kappa h_{\mu\nu}$, where the perturbation $h_{\mu\nu}$ is identified with the graviton.

The heavy-field limit of a real scalar field can be expressed in terms of a complex scalar field $\chi$ by employing a suitable field-redefinition. Motivated by earlier analyses in Refs.~\cite{Guth:2014hsa,Braaten:2018lmj,NRRealScalar}, we decompose
\begin{align}
    \phi\rightarrow\frac{1}{\sqrt{2m}}\left(e^{-imv\cdot x}\chi+e^{imv\cdot x}\chi^{*}\right).
\end{align}
Substituting this into Eq.~\eqref{eqn:ScGravLag} and dropping quickly oscillating terms (those proportional to $e^{\pm2imv\cdot x}$) gives the HBET Lagrangian for scalars:
\begin{align}
	\sqrt{-g}\mathcal{L}_{\text{HBET}}^{s=0}&=\sqrt{-g}\chi^{*}\left[g^{\mu\nu}iv_{\mu}\partial_{\nu}+\frac{1}{2}m(g^{\mu\nu}v_{\mu}v_{\nu}-1)-\frac{1}{2m}g^{\mu\nu}\partial_{\mu}\partial_{\nu}\right]\chi + \mathcal{O}(1/m^2).\label{eq:HSBET}
\end{align}

Comparing the Feynman rules for this theory in Appendix \ref{sec:FeynRules} with the Feynman rules for the full theory in Ref.~\cite{Holstein:2008sx}, we see that they are related by simply decomposing the momenta as in Eq.~\eqref{eqn:HQETMomDecomposition} and dividing by $2m$.

Next, we consider the case of a heavy spin-1/2 particle. We begin with the Lagrangian of a minimally coupled Dirac field $\psi$
\begin{align}
	\sqrt{-g} \mathcal{L}_{\text{grav}}&=\sqrt{-g}\bar{\psi}\left(i{e^{\mu}}_{a}\gamma^{a}\,D_{\mu}-m\right)\psi\label{eq:SpinorGravLag},
\end{align}
where ${e^{\mu}}_{a}$ is a vierbein, connecting curved space (with Greek indices) and flat space (with Latin indices) tensors. The expansion of the vierbein in terms of the metric perturbation is given in Ref.~\cite{Holstein:2008sx}. The covariant derivative is \cite{VierGR}
\begin{align}
D_{\mu}\psi \equiv\left(\partial_{\mu}+\frac{i}{2}{\omega_{\mu}}^{ab}_{\,}\sigma_{ab}\right)\psi,
\end{align}
where the spin connection ${\omega_{\mu}}^{ab}_{\,}$ is given in terms of vierbeins in Eq.~(41) of Ref.~\cite{VierGR}. To quadratic order in the graviton field, the spin-connection is \cite{Holstein:2008sx}
\begin{align}
    {\omega_{\mu}}^{ab}_{\,}&=-\frac{\kappa}{4}\partial^{b}{h_{\mu}}^{a}-\frac{\kappa^{2}}{16}h^{\rho b}\partial_{\mu}{h^{a}}_{\rho}+\frac{\kappa^{2}}{8}h^{\rho b}\partial_{\rho}{h_{\mu}}^{a}-\frac{\kappa^{2}}{8}h^{\rho b}\partial^{a}{h_{\mu}}^{\rho}-(a\leftrightarrow b).\label{eq:SpinCon}
\end{align}
Eq.~\eqref{eq:SpinCon} differs from that in Ref.~\cite{Holstein:2008sx} by a factor of $-1/2$. The spin connection of Ref.~\cite{VierGR} differs from that of Ref.~\cite{Holstein:2008sx} by this same factor, and we use the connection of Ref.~\cite{VierGR}. % as their Lagrangian is the same as our starting Lagrangian in Eq.~\eqref{eq:SpinorGravLag}.

We make the same decomposition of the fermion field $\psi$ as in HQET, 
Eq.~\eqref{eqn:HQETSpinor}, and integrate out the anti-field by substituting its equation of motion.
As for HQET, the HBET Lagrangian has a non-local form:
\begin{align}
	\label{eq:HBET}
	\sqrt{-g}\mathcal{L}^{s=1/2}_{\rm HBET} =& 
	\sqrt{-g} \bar Q \left[
		(i \slashed \nabla + \mathcal{B} )
		+
		(i \slashed \nabla + \mathcal{B} )
		P_{-} \frac{1}{2m - (i \slashed \nabla + \mathcal{B} )P_{-}}
		(i \slashed \nabla + \mathcal{B} )
	\right] Q ,
\end{align}
where $\slashed \nabla \equiv \delta^{\mu}_{a} \gamma^{a} \nabla_{\mu}$ and
\begin{align}
	\mathcal{B} = ( e^{\mu}_{a} - \delta^{\mu}_{a} ) ( i \gamma^{a} \nabla_{\mu} + m \gamma^{a} v_{\mu} ) .
\end{align}
This is the main result of the paper.

We can recover a local form of this Lagrangian by expanding the denominator in both $1/m$ and $\kappa$. We will only need vertices involving two spinors and at most two gravitons, so we expand up to $\mathcal{O}(\kappa^{2})$. The resulting Feynman rules are given in Appendix~\ref{sec:FeynRules} for reference.

Although we started with massive matter fields, Eqs.~\eqref{eq:HSBET} and \eqref{eq:HBET} contain no mass terms for the matter fields. The propagating modes of HBET are therefore massless, so their momenta scale with $\hbar$ in the classical limit. As in the case of HQET, this allows us to interpret the operator expansion of HBET as an expansion in $\hbar$.

The Feynman rules of both theories (Appendix~\ref{sec:FeynRules}) are suggestive of the universality of the multipole expansion from Ref.~\cite{Holstein:2008sx}; all terms present in the scalar Feynman rules also appear in the spinor Feynman rules. There are, of course, extra terms in the spinor Feynman rules which encode spin effects. Moreover, we find additional spin-independent terms in the spinor Feynman rules that do not appear in the scalar rules. 
This is not necessarily inconsistent with Ref.~\cite{Holstein:2008sx}: as will be discussed further below, we expect these additional terms to not contribute to the
properly defined potential at one-loop level.% as the universality presented there is at the level of the non-relativistic amplitude.

\section{Long range \texorpdfstring{$2\rightarrow2$}{2-to-2} gravitational scattering amplitudes}
\label{sec:Amps}

We will demonstrate the utility of the above EFTs for systems of two heavy particles. We do so by calculating the amplitudes for the scattering of scalars and fermions mediated by gravitons up to the leading quantum order at one-loop level. To maximize the efficiency of the computation of the following amplitudes, one could obtain them as double copies of HQET amplitudes. Focusing on the validation of HBET, however, we compute them using standard Feynman diagram techniques applied directly to the HBET Lagrangians in Eqs.~\eqref{eq:HSBET} and \eqref{eq:HBET}, with graviton dynamics described by the usual Einstein-Hilbert action,
\begin{align}
    S_{\text{GR}}&=\frac{1}{16\pi G}\int d^{4}x\sqrt{-g}R.
\end{align}
To obtain the classical portions of the amplitudes, we use only the HBET operators described in Section~\ref{sec:hbarCount}. The leading quantum terms arise by also including operators that scale with one more factor of $\hbar$.

\iffalse
Loop integrals contain both classical and quantum pieces.
We therefore expand all loop integrals in $|q|=\hbar|\overline{q}|$. In both tree and loop amplitudes, we only keep terms up to $\mathcal{O}(|q|^{0})$ for the spinless, $\mathcal{O}(|q|^{1})$ for the spin-orbit, and $\mathcal{O}(|q|^{2})$ for the spin-spin contributions, as these are the combinations up to the leading quantum order in $\hbar$. Higher order terms in the $|q|$ expansion are subleading quantum mechanical contributions.
\fi

In what follows we make use of the reparameterization invariance of HBET \cite{Luke:1992cs,CHEN1993421,Finkemeier:1997re} to work in a frame in which the initial momenta are $p_{i}^{\mu}=m_{i}v_{i}^{\mu}$, where $v_{i}^{\mu}$ is the initial four-velocity of particle $i$. We then define $\omega\equiv v_{1\mu}v_{2}^{\mu}$, which, in such a frame, is related to the Mandelstam variable $s=(p_1+p_2)^2$ via
\begin{align}
    s-s_{0}&=2m_{1}m_{2}(\omega-1)\label{eqn:NRCondition},
\end{align}
where $s_{0}\equiv(m_{1}+m_{2})^{2}$. From Eq.~\eqref{eqn:NRCondition} it is evident that the non-relativistic limit of the kinematics of both particles, $s-s_{0}\rightarrow0$, is equivalent to the limit $\omega\rightarrow 1$. As a check on the results, we reproduce the amplitudes in Ref.~\cite{Holstein:2008sx} in the non-relativistic limit.

%\subsection{Post-Minkowskian gravity}

Amplitudes for scalar-scalar scattering arise as a portion of the fermion-fermion scattering amplitude \cite{Holstein:2008sx}. For this reason we present here the amplitudes for fermion-fermion scattering.

\subsection{First Post-Minkowskian Order}

\begin{figure}
    \centering
    \includegraphics[scale=1]{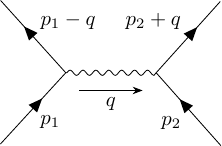}
    \caption[Tree-level scattering Feynman diagram.]{Classical scattering of two particles at tree-level.\label{fig:tChan}}
\end{figure}

At 1PM order, the relevant diagram is the tree-level graviton exchange diagram,
shown in Fig.~\ref{fig:tChan}. Using the $\hbar$-counting in Section~\ref{sec:hbarCount},
we see that the coupling constants provide one inverse power of $\hbar$, while the 
graviton propagator scales as $1/\hbar^2$. The leading tree-level amplitude becomes
%
\iffalse

We begin with the 1PM amplitude. There is only one diagram to consider; it is the tree-level graviton exchange diagram depicted in Fig.~\ref{fig:tChan}. Excluding the kinematic factors arising from the vertices, this diagram must scale with three inverse powers of $\hbar$: one comes from the two powers of $\sqrt{G}$ at the vertices, and two come from the momentum of the internal graviton. Thus, using only the first line of the Feynman rule in Eq.~\eqref{eq:HBET3pt}, we directly compute
\fi
\begin{align}
    \cM^{(1)}_{t}=-\frac{4\pi m_{1}m_{2}G}{\hbar^{3}q^{2}}&\left[(2\omega^{2}-1)\mathcal{U}_{1}\mathcal{U}_{2}+\frac{2i\omega}{m_{1}^{2}m_{2}}\mathcal{E}_{1}\mathcal{U}_{2}+\frac{2i\omega}{m_{1}m_{2}^{2}}\mathcal{E}_{2}\mathcal{U}_{1}\right. \notag \\
    &\quad\left.-\frac{1}{m_{1}^{3}m_{2}^{3}}\mathcal{E}_{1}\mathcal{E}_{2}+\frac{\omega}{m_{1}^{2}m_{2}^{2}}\mathcal{E}^{\mu}_{1}\mathcal{E}_{2\mu}\right].
\end{align}
This is in agreement with Ref.~\cite{Holstein:2008sx} at leading order in $\mathcal{O}(|q|)$. 
We use the shorthand notation 
\begin{subequations}
\begin{align}
    \mathcal{U}_{1}&\equiv\bar{u}(p_{1}-q)u(p_{1})\equiv\bar{u}_{2}u_{1},\\
    \mathcal{U}_{2}&\equiv\bar{u}(p_{2}+q)u(p_{2})\equiv\bar{u}_{4}u_{3},\\
    \mathcal{E}_{i}&\equiv\epsilon^{\mu\nu\alpha\beta}p_{1\mu}p_{2\nu}\bar{q}_{\alpha}S_{i\beta}, \\
    \mathcal{E}^{\mu}_{i}&\equiv\epsilon^{\mu\nu\alpha\beta}p_{i\nu}\bar{q}_{\alpha}S_{i\beta},
\end{align}
 with the relativistic normalization of the spinors, $\bar{u}(p)u(p)=2m$. The Levi-Civita tensor is defined by $\epsilon^{0123}=1$. 
The spin vector is defined as 
\begin{align}
    S_{i}^{\mu}\equiv\frac{1}{2}\bar{u}_{2i}\gamma_{5}\gamma^{\mu}u_{2i-1},
\end{align}
where $\gamma_{5}\equiv-i\gamma^{0}\gamma^{1}\gamma^{2}\gamma^{3}$. The definition of the HQET spinor in Eq.~\eqref{eq:HQETSpinor} automatically imposes the orthogonality of the spin vector and the momentum of the corresponding particle, since it implies the relation $\slashed{v}u=u$.
\end{subequations}

\begin{figure}
    \centering
    \begin{subfigure}[b]{0.25\textwidth}
    \includegraphics[width=\textwidth]{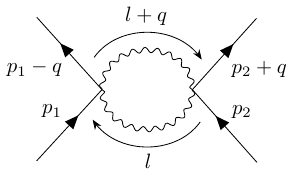}
    \caption{}
    \end{subfigure}\hspace{1em}
    \begin{subfigure}[b]{0.25\textwidth}
    \includegraphics[width=\textwidth]{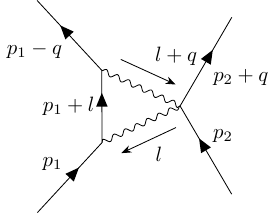}
    \caption{}
    \end{subfigure}
    \begin{subfigure}[b]{0.25\textwidth}
    \includegraphics[width=\textwidth]{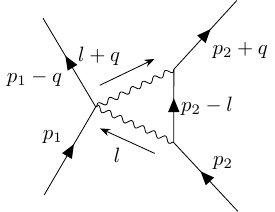}
    \caption{}
    \end{subfigure}\hspace{1em}
    \begin{subfigure}[b]{0.25\textwidth}
    \includegraphics[width=\textwidth]{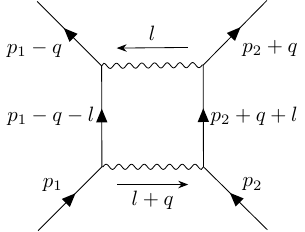}
    \caption{}
    \end{subfigure}\hspace{1em}
    \begin{subfigure}[b]{0.25\textwidth}
    \includegraphics[width=\textwidth]{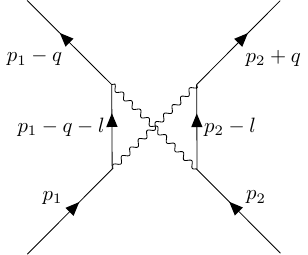}
    \caption{}
    \end{subfigure}
    \begin{subfigure}[b]{0.3\textwidth}
    \includegraphics[width=\textwidth]{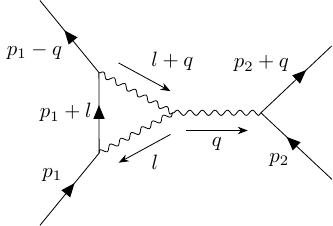}
    \caption{}
    \end{subfigure}\hspace{1em}
    \begin{subfigure}[b]{0.3\textwidth}
    \includegraphics[width=\textwidth]{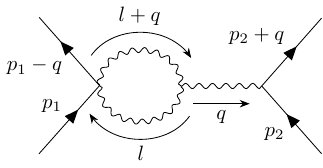}
    \caption{}
    \end{subfigure}
    \begin{subfigure}[b]{0.3\textwidth}
    \includegraphics[width=\textwidth]{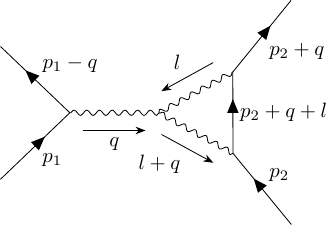}
    \caption{}
    \end{subfigure}\hspace{1em}
    \begin{subfigure}[b]{0.3\textwidth}
    \includegraphics[width=\textwidth]{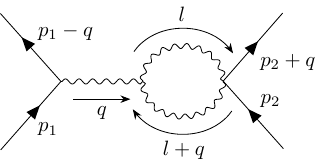}
    \caption{}
    \end{subfigure}\hspace{1em}
     \begin{subfigure}[b]{0.31\textwidth}
    \includegraphics[width=\textwidth]{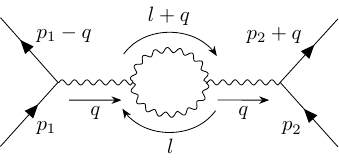}
    \caption{}
    \end{subfigure}\hspace{1em}
     \begin{subfigure}[b]{0.31\textwidth}
    \includegraphics[width=\textwidth]{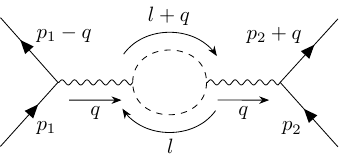}
    \caption{}
    \end{subfigure}\hspace{1em}
    \caption{The one-loop Feynman diagrams containing non-analytic pieces that contribute to the classical scattering of two particles in GR. Solid lines represent fermions, wavy lines represent gravitons, and dashed lines represent the ghost field arising from working in the harmonic gauge \cite{Holstein:2008sx}.}
    \label{fig:GR1Loop}
\end{figure}

\subsection{Second Post-Minkowskian Order}
At 2PM order, eleven one-loop diagrams can contribute, shown in Fig.~\ref{fig:GR1Loop}.
Only triangles or box diagrams contribute to the classical amplitude, but as we also compute
the leading quantum contributions, all eleven diagrams are needed.

\iffalse
Next we turn to the 2PM order, where there are eleven diagrams that contribute. They are shown in Fig.~\ref{fig:GR1Loop}. As we will be computing the amplitude up to the leading order quantum terms, we must consider all eleven diagrams. To obtain just the classical contributions, however, one would need only consider the six diagrams containing either triangles or boxes. %Then, in the triangles, we must use only the operators that contain no $\hbar$ dependence apart from in the coupling constant, whereas the boxes need operators with one positive power of $\hbar$ apart from the coupling constant.
\fi

For clarity, we split the 2PM amplitude into three parts: the spinless, spin-orbit, and spin-spin contributions. These are, respectively,
\begin{subequations}
{\small
\begin{align}
   &\cM^{(2)}_{\text{spinless}}=\nonumber \\ &\quad\frac{G^{2}}{\hbar^{3}}m_{1}m_{2}\mathcal{U}_{1}\mathcal{U}_{2}S\frac{3}{2}(5\omega^{2}-1)(m_{1}+m_{2})\notag \\
    &\quad+\frac{G^{2}\mathcal{U}_{1}\mathcal{U}_{2}L}{30\hbar^{2}(\omega^{2}-1)^{2}}\left[2m_{1}m_{2}(18\omega^{6}-67\omega^{4}+50\omega^{2}-1)\right.\notag \\
    &\quad
    \left.-60m_{1}m_{2}\omega(12\omega^{4}-20\omega^{2}+7)L_{\times}(\omega)\sqrt{\omega^{2}-1}\right.%\notag \\
   % &\quad
    \left.-15i\pi(m_{1}^{2}+m_{2}^{2})(24\omega^{4}-37\omega^{2}+13)\sqrt{\omega^{2}-1}\right.\notag \\
    &\quad\left.-\frac{120}{\hbar^{2}\overline{q}^{2}}i\pi m_{1}^{2}m_{2}^{2}(4\omega^{6}-8\omega^{4}+5\omega^{2}-1)\sqrt{\omega^{2}-1}\right]\label{eq:2PMSless}, \\
    &\cM^{(2)}_{\text{spin-orbit}}=\nonumber\\&\quad \frac{G^{2}m_{1}m_{2}\omega(5\omega^{2}-3)S}{2\hbar^{3}(\omega^{2}-1)}\left[(3m_{1}+4m_{2})\frac{i\mathcal{U}_{1}\mathcal{E}_{2}}{m_{1}m_{2}^{2}}\right]\notag \\
    &\quad+\frac{G^{2}L}{10\hbar^{2}(\omega^{2}-1)^{2}}\left\{2m_{1}m_{2}\omega(\omega^{2}-1)(46\omega^{2}-31)-20m_{2}^{2}i\pi\omega(\omega^{2}-2)\sqrt{\omega^{2}-1}\right.\notag \\
    &\quad\left.-\frac{80}{\hbar^{2}\overline{q}^{2}}i\pi m_{1}^{2}m_{2}^{2}\omega(\omega^{2}-1)(2\omega^{2}-1)\sqrt{\omega^{2}-1}-5m_{1}^{2}i\pi\omega(12\omega^{4}-10\omega^{2}-5)\sqrt{\omega^{2}-1}\right.\notag \\
    &\quad\left.-5m_{1}m_{2}[(40\omega^{4}-48\omega^{2}+7)L_{\times}(\omega)-(8\omega^{4}-1)(L_{\square}(\omega)+i\pi)]\sqrt{\omega^{2}-1}\right\}\frac{i\mathcal{U}_{1}\mathcal{E}_{2}}{m_{1}m_{2}^{2}}+(1\leftrightarrow2)\label{eq:2PMSO}, \\
    &\cM^{(2)}_{\text{spin-spin}}=\nonumber \\&\quad G^{2}(m_{1}+m_{2})\frac{S}{\hbar^{3}}\left[\frac{(20\omega^{4}-21\omega^{2}+3)}{2(\omega^{2}-1)}(\bar{q}\cdot S_{1}\bar{q}\cdot S_{2}-\bar{q}^{2}S_{1}\cdot S_{2})+\frac{2\bar{q}^{2}\omega^{3}(5\omega^{2}-4)}{m_{1}m_{2}(\omega^{2}-1)^{2}}p_{2}\cdot S_{1}p_{1}\cdot S_{2}\right]\notag \\
    &\quad+\frac{G^{2}L}{\hbar^{3}m_{1}m_{2}}[m_{1}C_{1}(m_{1},m_{2})p_{2}\cdot S_{1}\bar{q}\cdot S_{2}-m_{2}C_{1}(m_{2},m_{1})\bar{q}\cdot S_{1}p_{1}\cdot S_{2}]\notag \\
    &\quad+\frac{G^{2} L}{60m_{1}m_{2}\hbar^{2}(\omega^{2}-1)^2}(2C_{2}\bar{q}\cdot S_{1}\bar{q}\cdot S_{2}+C_{3}\bar{q}^{2}S_{1}\cdot S_{2})%\notag \\
   % &\quad
    +\frac{G^{2}\bar{q}^{2}L}{20\hbar^{4}m_{1}^{2}m_{2}^{2}(\omega^{2}-1)^{5/2}}C_{4}p_{2}\cdot S_{1}p_{1}\cdot S_{2}\label{eq:2PMSS},
\end{align}
}
where
{\small
\begin{align}
    L_{\square}(\omega)&\equiv\log\left|\frac{\omega-1-\sqrt{\omega^{2}-1}}{\omega-1+\sqrt{\omega^{2}-1}}\right|, \\
    L_{\times}(\omega)&\equiv\log\left|\frac{\omega+1+\sqrt{\omega^{2}-1}}{\omega+1-\sqrt{\omega^{2}-1}}\right|, \\
    C_{1}(m_{i},m_{j})&\equiv%-\frac{\bar{q}^{2}S}{16m_{i}m_{j}(\omega^{2}-1)^{2}}[m_{i}^{2}(22\omega^{6}+20\omega^{4}-22\omega^{2}-4)+m_{j}^{2}\omega(73\omega^{4}-70\omega^{2}+13)\notag \\
    %&\quad+4m_{i}m_{j}(\omega+1)^{2}(20\omega^{3}-15\omega^{2}-6\omega+3)]+
    \frac{(8\omega^{4}-8\omega^{2}+1)}{(\omega^{2}-1)^{3/2}}i\pi(m_{i}+\omega m_{j}), \\
    C_{2}&\equiv60m_{1}m_{2}\omega\left((L_{\square}(\omega)+i\pi) (4\omega^{4}-2\omega^{2}-1)+L_{\times}(\omega)(-8\omega^{4}+14 \omega^{2}-5)\right)\sqrt{
    \omega^{2}-1}\notag \\
    &\quad-\frac{120}{\hbar^{2}\overline{q}^{2}}i\pi m_{1}^{2}m_{2}^{2}(\omega^{2}-1)(1-2\omega ^{2})^{2}
   \sqrt{\omega^{2}-1}\notag \\
   &\quad-30i\pi(m_{1}^{2}+m_{2}^{2})(2\omega^{6}-4\omega^{4}- \omega^{2}+2)\sqrt{\omega^{2}-1}\notag \\
   &\quad+2m_{1}m_{2}(\omega^{2}-1)\left(258\omega^{4}-287\omega^{2}+29\right), \\
   C_{3}&\equiv60m_{1}m_{2}\omega  \left((L_{\square}(\omega)+i\pi)(3-4\omega^{2})+12
   L_{\times}(\omega)(2\omega^{4}-3\omega^{2}+1)\right)\sqrt{\omega^{2}-1}\notag \\
   &\quad+\frac{120}{\hbar^{2}\overline{q}^{2}}i\pi m_{1}^{2}m_{2}^{2}(\omega^{2}-1)(8\omega^{4}-8\omega^{2}+1)\sqrt{\omega^{2}-1}\notag \\
   &\quad+15i\pi(m_{1}^{2}+m_{2}^{2})(8\omega^{6}-12
   \omega^{4}-3\omega^{2}+5)\sqrt{\omega^{2}-1}\notag \\
   &\quad+4m_{1}m_{2}(\omega^{2}-1) \left(-258 \omega ^4+287 \omega
   ^2-44\right), \\
   C_{4}&\equiv%10 m_{1}m_{2} \left((L_{\square}(\omega)+i\pi) \left(8\omega^{6}+16\omega^{4}-19\omega^{2}+1\right)+L_{\times}(\omega) \left(-40\omega^{6}+64\omega^{4}-25\omega^{2}+1\right)\right)\notag \\
   %&\quad+4 m_{1}m_{2}\omega\left(86\omega^{4}-47\omega^{2}-24\right)\sqrt{\omega ^2-1}\notag \\
   %&\quad
   -\frac{40}{\bar{q}^{2}}i\pi  m_{1}^2 m_{2}^2 \omega  \left(\omega ^2-1\right) \left(8 \omega ^4-8 \omega ^2+1\right)%\notag \\
   %&\quad-5i\pi\left(m_{1}^2+m_{2}^2\right)(8\omega^{6}-28\omega^{4}+5\omega^{2}+9)
   .
\end{align}
}
\end{subequations}

The classical contributions are in the first lines of each of Eqs.~\eqref{eq:2PMSless}-\eqref{eq:2PMSO}, and in the first and second lines of Eq.~\eqref{eq:2PMSS}. The classical spinless contribution is in agreement with Ref.~\cite{Cheung:2018wkq}.
The classical spin-orbit contribution is consistent with the spin holonomy map of Ref.~\cite{Bini:2018ywr}. 
The classical spin-spin contribution compliments the results in Ref.~\cite{Chung:2018kqs}. 
In particular, we find that the coefficient of $(-q\cdot S_{1}q\cdot S_{2})$ in Eq.~\eqref{eq:2PMSS} agrees with $A^{2\text{ PM}}_{1,1}$ in Eq.~(7.18) in Ref.~\cite{Chung:2018kqs}, which is the corresponding coefficient in the Leading Singularity approach \cite{Cachazo:2017jef,Guevara:2017csg}, whereas the remainder of the terms are not presented therein.
To the best of our knowledge, this is the first presentation of the leading quantum contributions 
to the spinless, spin-orbit and spin-spin amplitudes at 2PM order. There are additional spin-spin terms at the quantum level proportional to $p_{i}\cdot S_{j}$ for $i\neq j$ that we have not included in our calculation.
%The 2PM classical amplitudes for the spinless, spin-orbit and spin-spin contributions
%compliment results in the literature \cite{Cheung:2018wkq,Bini:2018ywr,Chung:2018kqs}. 
We note that additional spin
quadrupole terms are also present at the second order in spin, which can be calculated
from vector-scalar scattering.

To obtain the results in this section we have made use of the identity
\begin{align}
    \bar{u}_{2i}\sigma^{\mu\nu}u_{2i-1}=-2\epsilon^{\mu\nu\alpha\beta}v_{i\alpha}S_{i\beta},\label{eqn:HeavyGordDecomp}
\end{align}
which is valid for HQET spinors. This identity merits some discussion. Replacing the HQET spinors by Dirac spinors (denoted with a subscript D), the identity becomes
\begin{align}
    \bar{u}_{2i,\text{D}}\sigma^{\mu\nu}u_{2i-1,\text{D}}=-2ip_{2i-1}^{[\nu}\bar{u}_{2i,\text{D}}\gamma^{\mu]}u_{2i-1,\text{D}}-\frac{2}{m}\epsilon^{\mu\nu\alpha\beta}p_{2i-1,\alpha}S_{i,\text{D}\beta}.\label{eqn:DiracGordDecomp}
\end{align}
The second term above is the same as in Eq.~\eqref{eqn:HeavyGordDecomp}. The first, by contrast, arises only with Dirac spinors, and through the Gordon decomposition contains both a spinless term involving only the spinor product $\mathcal{U}_{i}$, and a term like that on the left hand side of the equation. Eq.~\eqref{eqn:DiracGordDecomp} thereby mixes spinless and spin-inclusive effects. This is an advantage of this EFT approach, at least at one-loop level. Eq.~\eqref{eqn:HeavyGordDecomp} allows one to target spinless or spin-inclusive terms in the amplitude simply by ignoring or including operators involving the Dirac sigma matrix. It is also consistent with the universality of the spin-multipole expansion observed in Refs.~\cite{Holstein:2008sw,Holstein:2008sx}, where spin effects were found to not mix with, and to be corrections to the universal spin-independent amplitude.

At face value, there is one complication to this interpretation of Eq.~\eqref{eqn:HeavyGordDecomp}. Due to the heavy propagators, terms such as $\overline{u}\sigma^{\mu\nu}(1+\slashed{v})\sigma^{\alpha\beta}u$ begin to arise at one-loop level. Through some gamma matrix manipulations, it can be shown that these terms contain spinless (containing no sigma matrices) and spin-inclusive (containing one sigma matrix) components. At one-loop level the spinless components contribute to the classical and leading quantum portions of the spinless part of the amplitude only through the term proportional to $(m_{1}^{2}+m_{2}^{2})L$. As this term is purely imaginary, we expect it to be subtracted by the Born iteration when extracting the potential. Thus, if one is interested only in non-imaginary terms at one-loop% surviving the extraction of the potential at one-loop
, spinless or spin-inclusive terms can be independently targeted by exploiting the separation of spin effects at the level of the Lagrangian.

While spinless and spin-inclusive effects are cleanly separated in spinor HBET, the presence of additional spin-independent operators in spinor HBET compared to scalar HBET makes it ostensibly possible that the spinless parts of its amplitudes differ from the amplitudes of scalar HBET. In fact, calculating scalar-scalar scattering explicitly with scalar HBET, we find that the term proportional to $(m_{1}^{2}+m_{2}^{2})L$ in Eq.~\eqref{eq:2PMSless} does not arise. In addition to receiving contributions from the $\overline{u}\sigma^{\mu\nu}(1+\slashed{v})\sigma^{\alpha\beta}u$ tensor structure in the loop amplitudes --- a structure that certainly does not arise in scalar HBET --- it is also the only term that is affected by the spin-independent operators in spinor HBET that are not present in scalar HBET. We therefore find that we preserve the universality of the multipole expansion from Ref.~\cite{Holstein:2008sx} in the one-loop relativistic regime as well, up to terms which are subtracted by the Born iteration.

As a check on the validity of our results, we compare their non-relativistic limits with what exists in the literature, simply by taking the limit $\omega-1\rightarrow0$ in the PM amplitudes.
At 1PM order we find that our results agree with those in Ref.~\cite{Holstein:2008sx}. At 2PM the amplitudes above contain those in Ref.~\cite{Holstein:2008sx}, but there are two discrepancies:
\begin{enumerate}
    \item We find an additional spinless term that we expect to be subtracted by Born iteration, arising from the imaginary term proportional to $(m_{1}^{2}+m_{2}^{2})L$.
    \item The contraction $p_{i}\cdot S_{j}$ for $i\neq j$ vanishes in the non-relativistic limit. However, these terms in Eq.~\eqref{eq:2PMSS} also have denominators that vanish in this limit. Without knowing explicitly how $p_{i}\cdot S_{j}\rightarrow0$, we therefore cannot say that these terms will not remain in the limit.
\end{enumerate}
We note that this limit only represents the non-relativistic limit of the kinematics; the non-relativistic limit of the spinors must also be taken in order to obtain the fully non-relativistic amplitude.

\section{Conclusion}
\label{sec:Conclusions}

While significant progress has been made in understanding the relationship between gravitational scattering amplitudes and classical gravitational quantities, it remains uneconomical to extract the few classically contributing terms from the multitude of other terms that constitute the full amplitude. With an eye to addressing this inefficiency, we have introduced HBET, an EFT which describes the interactions of heavy scalars and heavy fermions with gravity. By restoring $\hbar$ at the level of the Feynman rules, we have been able to infer the $\hbar$-scaling of HBET operators, and exploit it to determine which operators can contribute to the classical amplitude at arbitrary loop order. One may see
the present construction as a step towards isolating just those terms of the
scattering amplitude that will contribute to the classical scattering of two massive objects,
order by order in the loop expansion. 
Crucially, a method does not yet exist to convert fully relativistic amplitudes including spin to interaction potentials.

We used HBET to directly calculate the 2PM classical gravitational scattering amplitude for the scattering of two fermions, and checked that the spinless part of the amplitude matches the amplitude for scalar-scalar scattering, up to terms that we expect to be subtracted from the potential. To validate the EFT, we compared the fully relativistic amplitudes and their non-relativistic limits with what has been previously calculated, and found agreement. 
We presented the classical and leading quantum spinless, spin-orbit and spin-spin contributions
at 2PM order, up to terms proportional to $p_{i}\cdot S_{j}$ where $i\neq j$ for the spin-spin contribution, complementing and extending the results in the literature.
%We presented for the first time the leading quantum spinless contributions at 2PM order, as well as the classical and leading quantum spin-orbit and spin-spin contributions at 2PM order.

While we derived HBET only for heavy particles of spin $s\leq1/2$, we believe it is possible to also derive an HBET for heavy higher-spin particles: as long as a Lagrangian can be written for a massive particle of spin $s$, %that is exact in that particle's mass, 
we can apply similar techniques to those herein to derive the HBET applicable for spin $s$. This would allow the computation of the classical amplitude for higher order terms in the multipole expansion. 
%One obvious drawback of this method is that terms in the spin multipole expansion with more than four powers of spin cannot be computed from HBET, due to the lack of a Lagrangian description of fundamental particles with $s>2$.

For full efficiency, the HBET formalism should be used in combination with modern scattering amplitude
techniques. First, the Feynman rules of scalar and spinor HBET, and the property in Eq.~\eqref{eqn:HeavyGordDecomp}, are suggestive of the universality of the multipole expansion presented in Ref.~\cite{Holstein:2008sx}. An interesting next step is to express the degrees of freedom of HBET in terms of massive on-shell variables \cite{Arkani-Hamed:2017jhn}. It would be interesting to study whether this universality can be made manifest in such variables, and how the observed separation of spinless and spin-inclusive effects arises. An on-shell formulation of HBET should also include an explicit $\hbar$ expansion, further elucidating the classical limit for amplitudes computed using on-shell variables. Moreover, the work in Refs.~\cite{Arkani-Hamed:2017jhn,Chung:2018kqs,Chung:2019duq} suggests that massive on-shell variables may facilitate the extension of HBET to higher spins. 
Second, as HQET is derived from QCD, and HBET is derived from GR, we expect the
double copy structure of the scattering amplitudes to still hold as a relation between the 
effective theories. While certainly not the only way to study such a relation, we expect it to be more readily apparent in on-shell versions of HQET and HBET. We leave the on-shell formulation of HBET for future work. %This relation would provide a new lens through which the classical double copy could be studied. We leave the details of these points for future work.

By combining the power counting (which
includes the $\hbar$ counting) and multipole expansion of the effective field theories with the on-shell formalism, unitarity methods, and the double copy,
we believe that higher order calculations are within reach.

\acknowledgments

We thank N. E. J. Bjerrum-Bohr,  Andrea Cristofoli, Aneesh Manohar, and Rafael Porto for useful discussions. We also thank Andr\`{e}s Luna for pointing out a typo in the electromagnetic spin-orbit amplitude. This project has received funding from the European Union's Horizon 2020 research and innovation programme under the Marie Sk\l{}odowska-Curie grant agreement No. 764850 "SAGEX".
The work of AH was supported in part by the Danish National Research Foundation (DNRF91) and
the Carlsberg Foundation.

\appendix
\section{HQET Spinors}
\label{sec:HeavySpinors}

In this section, we make precise the external states of the HQET spinor field by expressing them in terms of the external states of the original Dirac spinor field $\psi$. To do so, we begin with the mode expansion of $\psi$:
\begin{align}
    \psi(x)=\int\frac{d^{3}p}{(2\pi)^{3}}\frac{1}{\sqrt{2E_{\mathbf{p}}}}\sum_{s}\left(a^{s}_{\mathbf{p}}u^{s}_{\text{D}}(p)e^{-ip\cdot x}+b^{s\dagger}_{\mathbf{p}}w^{s}_{\text{D}}(p)e^{ip\cdot x}\right),
\end{align}
where $\mathbf{p}$ represents the three-momentum, $E_{\mathbf{p}}=\sqrt{\mathbf{p}^{2}+m^{2}}$, $s$ is a spin index, and $a^{s}_{\mathbf{p}}$ and $b^{s\dagger}_{\mathbf{p}}$ are annihilation and creation operators for the particle and antiparticle respectively. We use the unconventional notation $w_{D}$ for the antiparticle spinor to differentiate it from the four-velocity. The spinors $u^{s}_{\text{D}}(p)$ and $w^{s}_{\text{D}}(p)$ satisfy the Dirac equation,
\begin{subequations}
\begin{align}
    (\slashed{p}-m)u^{s}_{\text{D}}(p)&=0, \\
    (\slashed{p}+m)w^{s}_{\text{D}}(p)&=0.
\end{align}
\end{subequations}
Recall the definition of the HQET spinor field $Q_{v}$,
\begin{align}
    Q_{v}&=e^{imv\cdot x}\frac{1+\slashed{v}}{2}\psi,
\end{align}
where $v^{\mu}$ is defined by the HQET momentum decomposition in Eq.~\eqref{eqn:HQETMomDecomposition}. The mode expansion for $Q_{v}$ is then
\begin{align}
    Q_{v}(x)=\int\frac{d^{3}k}{(2\pi)^{3}}\frac{1}{\sqrt{2E_{\mathbf{p}}}}\sum_{s}\left(a^{s}_{\mathbf{p}}\frac{1+\slashed{v}}{2}u^{s}_{\text{D}}(p)e^{-ik\cdot x}+b^{s\dagger}_{\mathbf{p}}\frac{1+\slashed{v}}{2}w^{s}_{\text{D}}(p)e^{i(2mv+k)\cdot x}\right).
\end{align}
After the decomposition in Eq.~\eqref{eqn:HQETMomDecomposition}, the Dirac equation can be rewritten as
\begin{subequations}
\begin{align}
    \slashed{v}u^{s}_{\text{D}}(p)&=\left(1-\frac{\slashed{k}}{m}\right)u^{s}_{\text{D}}(p), \\
    \slashed{v}w^{s}_{\text{D}}(p)&=-\left(1-\frac{\slashed{k}}{m}\right)w^{s}_{\text{D}}(p).
\end{align}
\end{subequations}
Using this in the mode expansion for $Q_{v}$ we find
\begin{subequations}
\begin{align}
    Q_{v}(x)=\int\frac{d^{3}k}{(2\pi)^{3}}\frac{1}{\sqrt{2E_{\mathbf{p}}}}\sum_{s}\left(a^{s}_{\mathbf{p}}u^{s}_{v}(p)e^{-ik\cdot x}+b^{s\dagger}_{\mathbf{p}}w^{s}_{v}(p)e^{i(2mv+k)\cdot x}\right)\label{eq:QModeDecomp}.
\end{align}
where
\begin{align}
    u^{s}_{v}(p)&\equiv\left(1-\frac{\slashed{k}}{2m}\right)u^{s}_{\text{D}}(p), \\
    w^{s}_{v}(p)&\equiv\frac{\slashed{k}}{2m}w^{s}_{\text{D}}(p).
\end{align}
\end{subequations}
Similarly, the mode expansion of $\tilde{Q}_{v}$ is
\begin{align}
    \tilde{Q}_{v}=\int\frac{d^{3}k}{(2\pi)^{3}}\frac{1}{\sqrt{2E_{\mathbf{p}}}}\sum_{s}\left(a^{s}_{\mathbf{p}}\frac{\slashed{k}}{2m}u^{s}_{D}(p)e^{-ik\cdot x}+b^{s\dagger}_{\mathbf{p}}\left(1-\frac{\slashed{k}}{2m}\right)w^{s}_{D}(p)e^{i(2mv+k)\cdot x}\right).
\end{align}

%These definitions of the external states of the field $Q$ illustrate that, when $k^{\mu}=0$ or $m=\infty$, $Q$ contains only the particle states of the field $\psi$. Furthermore, 
The mode expansion in Eq.~\eqref{eq:QModeDecomp} makes it apparent that, when considering only particles and not antiparticles, the derivative of $Q_{v}$ translates to a factor of the residual momentum $k^{\mu}$ in the Feynman rules.

\section{Heavy Scalar Effective Theory}
\label{sec:HSQED}

For completeness we include the derivation of an effective theory for Scalar Quantum Electrodynamics (SQED). 
That is, we want the effective theory that arises when $\phi$ in
\begin{align}
\mathcal{L}_{\text{SQED}}&=(D_{\mu}\phi)^{*}D^{\mu}\phi-m^{2}\phi^{2},\ \qquad  D^{\mu}\phi=(\partial^{\mu}+ieA^{\mu})\phi, \label{eqn:LagSQED}
\end{align}
is very massive. To do so, we simply make the field redefinition \cite{Schwartz:2013pla}
\begin{align}
\phi\rightarrow\frac{e^{-imv\cdot x}}{\sqrt{2m}}\left( \chi + \tilde\chi \right).
\label{eqn:HeavyScalar}
\end{align}
The anti-field $\tilde \chi$ is to be integrated out. At leading order, we can drop this term.
Inserting Eq.~\eqref{eqn:HeavyScalar} into Eq.~\eqref{eqn:LagSQED}, and performing a field redefinition to eliminate redundant operators, we obtain Heavy Scalar Effective Theory (HSET):
\begin{align}
	\mathcal{L}_{\text{HSET}}&=\chi^{*}\left(iv\cdot D-\frac{D^{2}}{2m}\right)\chi + \mathcal{O}(1/m^3).
\end{align}
Higher order terms can be restored by keeping contributions coming from integrating out the 
anti-field.

\section{Long range \texorpdfstring{$2\rightarrow2$}{2-to-2} electromagnetic scattering amplitudes}
\label{sec:AmpsEM}

In this section we demonstrate that HSET and HQET can be used to calculate the classical and leading quantum contributions to the $2\rightarrow2$ scattering amplitudes. We present here the results up to one-loop order. As in the gravity case, electromagnetic interactions also possess a universal spin-multipole expansion \cite{Holstein:2008sw}, so we present this calculation using HQET.

At tree level, the diagram in Fig.~\ref{fig:tChan} is once again the only one that contributes. The amplitude is, up to leading order in $|q|$,
\begin{align}
    \cA^{(0)}&=\frac{4\pi\alpha}{\hbar^{3}q^{2}}\left[\omega\mathcal{U}_{1}\mathcal{U}_{2}+\frac{i\mathcal{U}_{1}\mathcal{E}_{2}}{m_{1}m_{2}^{2}}+\frac{i\mathcal{E}_{1}\mathcal{U}_{2}}{m_{1}^{2}m_{2}}+\frac{\omega}{m_{1}m_{2}}\mathcal{E}^{\mu}_{1}\mathcal{E}_{2\mu}\right].
\end{align}
This amplitude is in agreement with Ref.~\cite{Holstein:2008sw} in the relativistic and non-relativistic regimes.

At one-loop level, the abelian nature of QED reduces the number of relevant diagrams compared to the gravity case. There are only five relevant diagrams in the electromagnetic case: they are diagrams (a) to (e) in Fig.~\ref{fig:GR1Loop}. Of course, the wavy lines are reinterpreted as photons. We find the amplitude
\begin{subequations}
{\small
\begin{align}
   \cA^{(1)}_{\text{spinless}}&=\frac{\alpha^{2}}{\hbar^{3}m_{1}m_{2}}\left[S(m_{1}+m_{2})-\frac{\hbar L}{2m_{1}m_{2}(\omega^{2}-1)^{2}}\left(2m_{1}m_{2}(\omega^{4}-1)\notag \right.\right. \\
    &\quad\left.\left.+4m_{1}m_{2}\omega(\omega^{2}-2)L_{\times}(\omega)\sqrt{\omega^{2}-1}\right.\right.\notag \\
    &\quad\left.\left.+(m_{1}^{2}+m_{2}^{2})i\pi(\omega^{2}-1)^{2}\sqrt{\omega^{2}-1}\right.\right.\notag \\
    &\quad\left.\left.+\frac{8i\pi}{\hbar^{2}\overline{q}^{2}}m_{1}^{2}m_{2}^{2}\omega^{2}(\omega^{2}-1)\sqrt{\omega^{2}-1}\right)\right]\mathcal{U}_{1}\mathcal{U}_{2}\label{eq:2PMSlessEM}, \\
   \cA^{(1)}_{\text{spin-orbit}}&=\frac{\alpha^{2}}{\hbar^{3}m_{1}m_{2}(\omega^{2}-1)}S \omega (m_{1} + 2m_{2})\frac{i\mathcal{U}_{1}\mathcal{E}_{2}}{m_{1}m_{2}^{2}}\notag \\
	%&\quad+\frac{\alpha^{2}L}{12\hbar^{2}m_{1}^{2}m_{2}^{2}(\omega^{2}-1)}
    %\left[12 m_{2}^{2}\sqrt{\omega^{2}-1}\omega(i\pi)\right.\notag \\
	%			&\quad\left.+2m_{1}m_{2}\sqrt{\omega^{2}-1}(9 \omega^{2} ( L_{\square}(\omega)+i\pi)+(9 - 6\omega^{2})L_{\times}(\omega) + \sqrt{\omega^{2}-1}(7\omega^{3} - 4\omega^{2} - \omega + 4))\right.\notag \\
	%			&\quad\left.+m_{1}^{2}\omega\sqrt{\omega^{2}-1}(-3(\omega^{2}-3)(L_{\square}(\omega)+i\pi)-6(\omega^{2}-2)L_{\times}(\omega)
	%				\right.\notag\\ &\left. \qquad\qquad
	%			+ \sqrt{\omega^{2}-1}(-\omega^{3} + 4\omega^{2} + \omega - 4))\right]\frac{i\mathcal{U}_{1}\mathcal{E}_{2}}{m_{1}m_{2}^{2}}\notag \\
    &\quad+\frac{\alpha^{2}L}{2\hbar^{2}m_{1}^{2}m_{2}^{2}(\omega^{2}-1)^{2}}
    \left[-m_{2}^{2}\omega(\omega^{2}-3)\sqrt{\omega^{2}-1}(i\pi)\right.\notag \\
				&\quad\left.+m_{1}m_{2}\sqrt{\omega^{2}-1}((2 \omega^{2}+1) ( L_{\square}(\omega)+i\pi)+(3 - 2\omega^{2})L_{\times}(\omega) + \sqrt{\omega^{2}-1}(\omega^{2} + 2\omega-1))\right.\notag \\
				&\quad\left.-m_{1}^{2}\omega(2\omega^{2}-3)\sqrt{\omega^{2}-1}(2\omega^{2}-3)(i\pi)\right]\frac{i\mathcal{U}_{1}\mathcal{E}_{2}}{m_{1}m_{2}^{2}}\notag \\
    &\quad-\frac{4i\pi\alpha^{2}L\omega}{\hbar^{4}\overline{q}^{2}\sqrt{\omega^{2}-1}}\frac{i\mathcal{U}_{1}\mathcal{E}_{2}}{m_{1}m_{2}^{2}}+\,(1\leftrightarrow2), \\
    \cA^{(1)}_{\text{spin-spin}}&=\frac{\alpha^{2}S(m_{1}+m_{2})}{\hbar^{3}m_{1}^{2}m_{2}^{2}(\omega^{2}-1)}\left[(2\omega^{2}-1)(q\cdot S_{1}q\cdot S_{2}-q^{2}S_{1}\cdot S_{2})+\frac{2q^{2}\omega^{3}}{m_{1}m_{2}(\omega^{2}-1)}p_{2}\cdot S_{1}p_{1}\cdot S_{2}\right]\notag \\
    &\quad+\frac{\alpha^{2}}{\hbar^{2}m_{1}m_{2}}[m_{1}C_{1}^{\prime}(m_{1},m_{2})p_{2}\cdot S_{1}q\cdot S_{2}-m_{2}C_{1}^{\prime}(m_{2},m_{1})q\cdot S_{2}p_{1}\cdot S_{2}]\notag \\
    &\quad+\frac{\alpha^{2}L}{2\hbar^{2}m_{1}^{3}m_{2}^{3}(\omega^{2}-1)^{2}}(C^{\prime}_{2}q\cdot S_{1}q\cdot S_{2}+2C_{3}^{\prime}q^{2}S_{1}\cdot S_{2})\notag \\
    &\quad+\frac{\alpha^{2}L}{2m_{1}^{3}m_{2}^{3}(\omega^{2}-1)^{5/2}}C_{4}^{\prime}p_{2}\cdot S_{1}p_{1}\cdot S_{2},
\end{align}
}
where
{\small
\begin{align}
    C_{1}^{\prime}(m_{i},m_{j})&\equiv-\frac{q^{2}S}{m_{i}^{3}m_{j}^{3}(\omega^{2}-1)^{2}}[m_{i}^{2}(3\omega^{4}+8\omega^{2}-3)+4m_{i}m_{j}(\omega+1)^{2}(2\omega-1)+2m_{j}^{2}\omega(5\omega^{2}-1)]\notag \\
    &\quad+\frac{L(2\omega^{2}-1)}{\hbar m_{i}^{2}m_{j}^{2}}i\pi(m_{i}+\omega m_{j}), \\
    %%%%%%%%%%%%%%%%
    C_{2}^{\prime}&\equiv4m_{1}m_{2}\omega
   \left(L_{\times}(\omega)+L_{\square}(\omega)\omega^{2}+i\pi\omega^{2}\right) \sqrt{\omega^{2}-1}
  \nonumber \\ 
   &\quad-i\pi(m_{1}^{2}+m_{2}^{2})\left(2\omega^{4}-5\omega^{2}+1\right)
   \sqrt{\omega^{2}-1}\notag \\
   &\quad+6m_{1}m_{2}\left(\omega ^2-1\right)^{2}-\frac{8}{\hbar^{2}\bar{q}^{2}}i\pi m_{1}^{2}m_{2}^{2}\omega^{2}\left(\omega^{2}-1\right)\sqrt{\omega^{2}-1}, \\
   %%%%%%%%%%%%
    C_{3}^{\prime}&\equiv2m_{1}m_{2}\omega  \left(-L_{\square}(\omega)+2
   L_{\times}(\omega) \left(\omega ^2-1\right)-i \pi \right) \sqrt{\omega^{2}-1}
   \nonumber\\
   &\quad+i\pi(m_{1}^{2}+m_{2}^{2})\left(2 \omega^{4}-4\omega^{2}+1\right)\sqrt{\omega^{2}-1}\notag \\
   &\quad+2m_{1}m_{2}\left(\omega ^2-1\right)\left(2-3 \omega ^2\right)+\frac{4}{\hbar^{2}\bar{q}^{2}}i\pi m_{1}^{2}m_{2}^{2}\left(\omega ^2-1\right)\left(2 \omega ^2-1\right)\sqrt{\omega^{2}-1}, \\
   %%%%%%%%%%%%%
   C_{4}^{\prime}&\equiv-\frac{4}{\hbar^{2}\bar{q}^{2}}i\pi m_{1}^{2}m_{2}^{2}\omega\left(\omega ^2-1\right)\left(2 \omega ^2-1\right)+6m_{1}m_{2}\omega ^3 \sqrt{\omega ^2-1}\notag \\
   &\quad+m_{1}m_{2} \left((L_{\square}(\omega)+i\pi)\left(2 \omega ^4+5 \omega
   ^2-1\right)+L_{\times}(\omega) \left(-2 \omega ^4+3 \omega ^2-1\right)\right)\notag \\
   &\quad-i \pi\omega \left(2 \omega ^4-6 \omega
   ^2+1\right) \left(m_{1}^2+m_{2}^2\right).
\end{align}
}
\end{subequations}
The non-relativistic limit of this amplitude is in agreement with Ref.~\cite{Holstein:2008sw}, with discrepancy number 2 from the gravitational case applying here as well.

Calculating explicitly the amplitude for scalar-scalar scattering using HSET, we find the same amplitude as in Eq.~\eqref{eq:2PMSlessEM}, but without the imaginary term proportional to $(m_{1}^{2}+m_{2}^{2})L$. This term vanishes in the non-relativistic limit, thus preserving the non-relativistic universality of the multipole expansion in Ref.~\cite{Holstein:2008sw}.
Furthermore, we expect it to be subtracted by the Born iteration when calculating the potential, thus extending the multipole univerality to the relativistic potential.

\section{Feynman rules}
\label{sec:FeynRules}

We list here the Feynman rules used to perform the calculations in this paper. 
Below we denote the matter wave vector entering the vertex by $k_{1}$ and the matter wave vector leaving by $k_{2}$. $q_{1}$ and $q_{2}$ are incoming photon (graviton) wave vectors with indices $\mu,\nu\, (\mu\nu,\alpha\beta)$, respectively. 

We use the photon propagator in the Feynman gauge. The graviton propagator, three graviton vertex, as well as the ghost propagator and two-ghost-one-graviton vertex are given in the harmonic gauge in Ref.~\cite{Holstein:2008sx}.

\subsection{Abelian HSET}
Starting with HSET, the one- and two-photon vertex Feynman rules are
\begin{subequations}
\begin{align}
	\tau^{\mu}_{\chi\chi^{*}\gamma}(m,v,k_{1},k_{2})&=-\frac{ie}{\sqrt{\hbar}}\left[v^{\mu}+\frac{\hbar}{2m}(k_{1}^{\mu}+k_{2}^{\mu}) + \mathcal{O}\left(\frac{\hbar^3}{m^3}\right)\right], \\
     \tau^{\mu\nu}_{\chi\chi^{*}\gamma\gamma}(m,v,k_{1},k_{2})&=\frac{ie^{2}}{m\hbar}\left[\eta^{\mu\nu}+
     \mathcal{O}\left(\frac{\hbar^2}{m^2}\right)\right].
\end{align}
\end{subequations}

\subsection{Scalar HBET}
For HSBET the one- and two-graviton vertex Feynman rules are
\begin{subequations}
\begin{align}
    \tau^{\mu\nu}_{\chi\chi^{*}h}(m,v,k_{1},k_{2})&=-\frac{i\kappa}{2\sqrt{\hbar}}\left\{mv^{\mu}v^{\nu}-\frac{\hbar}{2}[\eta^{\mu\nu}v_{\rho}(k_{1}^{\rho}+k_{2}^{\rho})-v^{\mu}(k_{1}^{\nu}+k_{2}^{\nu})-v^{\nu}(k^{\mu}_{1}+k_{2}^{\mu})]\right.\notag \\
    &\qquad\left.+\frac{\hbar^{2}}{2m}\left[(k_{1}^{\mu}k_{2}^{\nu}+k_{2}^{\mu}k_{1}^{\nu})-\eta^{\mu\nu}k_{1\alpha}k_{2}^{\alpha}\right] + \mathcal{O}\left(\frac{\hbar^3}{m^2}\right)\right\}, \\
    \tau^{\mu\nu,\alpha\beta}_{\chi\chi^{*}hh}(m,v,k_{1},k_{2})&=\frac{i\kappa^{2}}{\hbar}\left\{mv_{\tau}v_{\lambda}\left[I^{\mu\nu,\tau\gamma}{I_{\gamma}}^{\lambda,\alpha\beta}-\frac{1}{4}(\eta^{\mu\nu}I^{\alpha\beta,\tau\lambda}+\eta^{\alpha\beta}I^{\mu\nu,\tau\lambda})\right]\right.\notag \\
    &\qquad \left.+\frac{\hbar}{4}\{-P^{\mu\nu,\alpha\beta}v_{\rho}(k_{1}^{\rho}+k_{2}^{\rho})-(\eta^{\mu\nu}I^{\alpha\beta,\tau\lambda}+\eta^{\alpha\beta}I^{\alpha\beta,\tau\lambda})v_{\tau}(k_{1\lambda}+k_{2\lambda})\right.\notag \\
    &\qquad\left.+2I^{\mu\alpha,\tau\gamma}{I_{\gamma}}^{\lambda,\nu\beta}[v_{\tau}(k_{1\lambda}+k_{2\lambda})+v_{\lambda}(k_{1\tau}+k_{2\tau})]\}\right.\notag \\
    &\qquad\left.+\frac{\hbar^{2}}{4m}[-P^{\mu\nu,\alpha\sigma}k_{1\rho}k_{2}^{\rho}-(\eta^{\mu\nu}I^{\alpha\beta,\tau\lambda}+\eta^{\alpha\beta}I^{\mu\nu,\tau\lambda})k_{1\tau}k_{2\lambda}\right.\notag \\
    &\qquad\left.+2I^{\mu\alpha,\tau\gamma}{I_{\gamma}}^{\lambda,\nu\beta}(k_{1\tau}k_{2\lambda}+k_{2\tau}k_{1\lambda})] + \mathcal{O}\left(\frac{\hbar^3}{m^2}\right)\right\},
\end{align}
where
\begin{align}
    P_{\mu\nu,\alpha\beta}&=\frac{1}{2}(\eta^{\mu\alpha}\eta^{\nu\beta}+\eta^{\mu\beta}\eta^{\nu\alpha}-\eta^{\mu\nu}\eta^{\alpha\beta}).
\end{align}
\end{subequations}

The propagator in both scalar theories is
\begin{align}
    D^{s=0}_{v}(k)&=\frac{i}{\hbar v\cdot k}.
\end{align}

\subsection{Abelian HQET}

The one- and two-photon Feynman rules in HQET are
\begin{subequations}
\begin{align}
	\tau^{\mu}_{\bar{Q}Q\gamma}(m,v,k_{1},k_{2})&=-\frac{ie}{\sqrt{\hbar}}\left\{v^{\mu}+\frac{i}{2m}\sigma^{\mu\nu}(k_{2\nu}-k_{1\nu})+\frac{\hbar}{2m}(k_{1}^{\mu}+k_{2}^{\mu})_{\perp}\right.\notag \\
						    &\quad\left.+\frac{i\hbar}{8m^{2}}v_{\rho}\sigma_{\alpha\beta}(k^{\alpha}_{1}+k^{\alpha}_{2})_{\perp}\left[(k_{2}^{\rho}-k_{1}^{\rho})\eta^{\mu\beta}-(k_{2}^{\beta}-k_{1}^{\beta})\eta^{\mu\rho}\right]\right.\notag \\
    &\quad\left.+\frac{\hbar^{2}}{8m^{2}}[v^{\mu}(k_{2}-k_{1})^{2}-v_{\rho}(k_{2}^{\rho}-k_{1}^{\rho})(k_{2}^{\mu}-k_{1}^{\mu})]\right.\notag \\
    &\quad\left.+\frac{i\hbar^{2}}{8m^{3}}(k_{1\perp}^{2}+k_{2\perp}^{2})\sigma^{\mu\rho}(k_{2\rho}-k_{1\rho}) + \mathcal{O}\left(\frac{\hbar^3}{m^3}\right)\right\}, \\
	\tau^{\mu\nu}_{\bar{Q}Q\gamma\gamma}(m,v,k_{1},k_{2})&=\frac{ie^{2}}{m\hbar}\left\{\eta^{\mu\nu}_{\perp}+\frac{i}{4m}[\sigma^{\mu\nu}v_{\rho}(q_{2}^{\rho}-q_{1}^{\rho}) + v^{\mu} v^{\nu} \sigma^{\lambda\rho} v_{\lambda}(q_{1\rho} + q_{2\rho}) \right. \notag\\ &\left. 
		- v^{\mu} v_{\lambda} \sigma^{\lambda\nu} (v\cdot q_{2}) 
		- v^{\nu} \sigma^{\mu\lambda} q_{2\lambda} 
		- v^{\nu} v_{\lambda} \sigma^{\lambda\mu} (v\cdot q_{1}) 
		- v^{\mu} \sigma^{\nu\lambda} q_{1\lambda} 
	]\right.\notag \\
							     &\quad\left.-\frac{i\hbar}{4m^{2}}(k_{2\rho}+k_{1\rho})\sigma_{\alpha\beta}(\eta^{\rho\mu}_{\perp}\eta^{\beta\nu}q_{1}^{\alpha}+\eta^{\rho\nu}_{\perp}\eta^{\beta\mu}q_{2}^{\alpha}) + \mathcal{O}\left(\frac{\hbar^2}{m^2}\right)\right\} ,
\end{align}
\end{subequations}
where $k^{\mu}_{\perp} = k^{\mu} - v^{\mu} (v\cdot k)$ and $\eta^{\mu\nu}_{\perp} = \eta^{\mu\nu} - v^{\mu} v^{\nu}$.

\subsection{Spinor HBET}

Finally, the one- and two-graviton Feynman rules in HBET are
\begin{subequations}
\begin{align}
    \tau^{\mu\nu}_{\bar{Q}Qh}(m,v,k_{1},k_{2})&=\frac{i\kappa}{2\sqrt{\hbar}}\left\{-mv^{\mu}v^{\nu}+\frac{i}{4}(v^{\mu}\sigma^{\rho\nu}+v^{\nu}\sigma^{\rho\mu})(k_{2\rho}-k_{1\rho})\right.\notag \\
    &\quad \left.+\frac{\hbar}{2}\left[v_{\alpha}(k_{1}^{\alpha}+k_{2}^{\alpha})\eta^{\mu\nu} -v^{\mu}(k_{1}^{\nu}+k_{2}^{\nu})-v^{\nu}(k_{1}^{\mu}+k_{2}^{\mu})\right.\right.\notag \\
    &\quad\left.\left.+3v^{\mu}(k_{2}^{\nu}-k_{1}^{\nu})+3v^{\nu}(k_{2}^{\mu}-k_{1}^{\mu})-6\eta^{\mu\nu}v_{\rho}(k_{2}^{\rho}-k_{1}^{\rho})\right]\right.\notag \\
    &\quad\left.+\frac{i\hbar}{4m}\left[(k_{2\rho}-k_{1\rho})(k_{1}^{\mu}\sigma^{\rho\nu}+k_{1}^{\nu}\sigma^{\rho\mu})-v^{\mu}v^{\nu}\sigma^{\rho\tau}k_{2\rho}k_{1\tau}\right.\right.\notag \\
    &\quad\left.\left.-\frac{1}{2}v_{\rho}(k_{2}^{\rho}k_{2\tau}-k_{1}^{\rho}k_{1\tau})(v^{\mu}\sigma^{\tau\nu}+v^{\nu}\sigma^{\tau\mu})\right.\right.\notag \\
    &\quad\left.\left.+(k_{2\rho}-k_{1\rho})(k_{2}^{\mu}-k_{1}^{\mu})\sigma^{\rho\nu}+(k_{2\rho}-k_{1\rho})(k_{2}^{\nu}-k_{1}^{\nu})\sigma^{\rho\mu}\right]\right.\notag \\
    &\quad\left.+\frac{\hbar^{2}}{2m}\left[-(k_{1}^{\mu}k_{2}^{\nu}+k_{1}^{\nu}k_{2}^{\mu})+\eta^{\mu\nu}k_{1}^{\rho}k_{2\rho}+\frac{1}{2}v^{\mu}v^{\nu}k_{1\rho}k_{2}^{\rho}\right.\right.\notag \\
    &\quad\left.\left.+\frac{1}{2}\eta^{\mu\nu}(k_{2\rho}-k_{1\rho})(k_{2}^{\rho}-k_{1}^{\rho})-\frac{1}{2}(k_{2}^{\mu}-k_{1}^{\mu})(k_{2}^{\nu}-k_{1}^{\nu})\right.\right.\notag \\
    &\quad\left.\left.+\frac{1}{4}v_{\rho}v^{\mu}(k_{2}^{\nu}k_{2}^{\rho}+k_{1}^{\nu}k_{1}^{\rho})+\frac{1}{4}v_{\rho}v^{\nu}(k_{2}^{\mu}k_{2}^{\rho}+k_{1}^{\mu}k_{1}^{\rho})\right]
    + \mathcal{O}\left(\frac{\hbar^2}{m^2}\right)\right\},\label{eq:HBET3pt} \\
    %%%%%%%%%5
    \tau^{\mu\nu,\alpha\beta}_{\bar{Q}Qhh}(m,v,k_{1},k_{2})&=\frac{i\kappa^{2}}{\hbar}\left\{mv_{\kappa}v_{\lambda}\left[I^{\mu\nu,\kappa\gamma}{I_{\gamma}}^{\lambda,\alpha\beta}-\frac{1}{4}(\eta^{\alpha\beta}I^{\mu\nu,\kappa\lambda}+\eta^{\mu\nu}I^{\alpha\beta,\kappa\lambda})\right]\right.\notag \\
    &\quad\left.-\frac{i}{16}\epsilon^{\lambda\rho\tau\delta}\gamma_{\delta}\gamma_{5}({I^{\mu\nu,\kappa}}_{\lambda}{I^{\alpha\beta,}}_{\tau\kappa}q_{2\rho}+{I^{\alpha\beta,\kappa}}_{\lambda}{I^{\mu\nu,}}_{\tau\kappa}q_{1\rho})\notag\right. \\
    &\quad\left.+\frac{i}{16}v_{\kappa}v_{\sigma}v_{\rho}\sigma_{\lambda\tau}[I^{\mu\nu,\kappa\lambda}I^{\alpha\beta,\sigma\tau}(q_{2\rho}+k_{1\rho})+I^{\alpha\beta,\kappa\lambda}I^{\mu\nu,\sigma\tau}(q_{1\rho}+k_{1\rho})]\right.\notag \\
    &\quad\left.-\frac{i}{8}v_{\kappa}\sigma_{\lambda\tau}(k_{1\sigma}-k_{2\sigma})(I^{\mu\nu,\kappa\lambda}I^{\alpha\beta,\sigma\tau}+I^{\alpha\beta,\kappa\lambda}I^{\mu\nu,\sigma\tau})\right.\notag \\
    &\quad\left.-\frac{3i}{16}v_{\kappa}\sigma_{\sigma\rho}(k_{1}^{\rho}-k_{2}^{\rho})(I^{\mu\nu,\kappa\tau}I^{\alpha\beta,\sigma\tau}+I^{\alpha\beta,\kappa\tau}I^{\mu\nu,\sigma\tau})\right.\notag \\
    &\quad\left.+\frac{i}{8}v_{\kappa}\sigma_{\lambda\rho}(k_{1}^{\rho}-k_{2}^{\rho})(\eta^{\mu\nu}I^{\alpha\beta,\kappa\lambda}+\eta^{\alpha\beta}I^{\mu\nu,\kappa\lambda})\right.\notag \\
    &\quad\left.+\frac{i}{16}v_{\kappa}\sigma_{\lambda\rho}(k_{1}^{\rho}-k_{2}^{\rho})(v^{\mu}v^{\nu}I^{\alpha\beta,\kappa\lambda}+v^{\alpha}v^{\beta}I^{\mu\nu,\kappa\lambda})\right.\notag \\
    &\quad\left.+\frac{i}{8}v_{\kappa}\sigma_{\lambda\rho}(\eta^{\mu\nu}I^{\alpha\beta,\kappa\lambda}q_{1}^{\rho}+\eta^{\alpha\beta}I^{\mu\nu,\kappa\lambda}q_{2}^{\rho})\right.\notag \\
    &\quad\left.+\frac{i}{8}v_{\rho}\sigma_{\lambda\tau}(I^{\mu\nu,\kappa\lambda}I^{\alpha\beta,\rho\tau}q_{1\kappa}+I^{\alpha\beta,\kappa\lambda}I^{\mu\nu,\rho\tau}q_{2\kappa}) + \mathcal{O}(\hbar)\right\}\label{eq:HBET4pt},
\end{align}
where
\begin{align}
    I^{\mu\nu,\alpha\beta}&=\frac{1}{2}(\eta^{\mu\alpha}\eta^{\nu\beta}+\eta^{\mu\beta}\eta^{\nu\alpha}).
\end{align}
\end{subequations}
Based on the $\hbar$ counting, there are additional terms that could contribute to the amplitude,
but we find that they contribute only at subleading quantum levels, and thus don't include them.

The propagator in both spinor theories is
\begin{align}
    D^{s=\frac{1}{2}}_{v}(k)&=\frac{i}{\hbar v\cdot k}\frac{1+\slashed{v}}{2}.
\end{align}

\section{One-loop integral basis}
\label{sec:Integrals}

In this section, we point out some subtleties that arise from the linear matter propagators characteristic of HQET/HBET. We first address the appearance of non-analytical contributions to loop integrals when using linear matter propagators instead of quadratic ones. Then we discuss how we circumvent the infamous pinch singularity of HQET.

\subsection{Non-analytic portions of loop integrals}

Consider, for example, the box integral with quadratic massive propagators:
\begin{align}
	I_{\text{quad}} = \int \frac{d^4 l}{(2\pi)^4} \frac{1}{l^2 (l+q)^2 \left[(p_1 - l - q)^2 - m_1^2 + i\epsilon\right]\left[(p_2 + l+q)^2 - m_2^2 + i\epsilon\right]}.\label{eq:QuadBox}
\end{align}
Letting the incoming momenta be $p_1^\mu = m_1 v_1^\mu$ and $p_2^\mu = m_2 v_2^\mu$, and making explicit the factors of $\hbar$ from the massless momenta,
\begin{align}
	&I_{\text{quad}} = \\
	&\int \frac{d^4 \overline l}{(2\pi)^4}
	\frac{1}{ \overline l^2 (\overline l + \overline q)^2
		\left[-2m_1 \hbar v_1 \cdot(\overline l + \overline q) 
		+ \hbar^2 (\overline l + \overline q)^2 + i \epsilon\right]
		\left[ 2m_2 \hbar v_2 \cdot (\overline l + \overline q)
		+ \hbar^2 (\overline l + \overline q)^2 + i \epsilon\right]}. \nonumber
\end{align}
Note that the massive propagators remain quadratic in the loop momentum. 

The box integral with the linear massive propagators of HQET/HBET takes the form
\begin{align}
	I_{\text{HQET}} = \int \frac{d^4 l}{(2\pi)^4} \frac{1}{l^2(l+q^2)\left[-v_1\cdot (l+q) + 
	i \epsilon\right]\left[v_2\cdot (l+q) + i\epsilon\right]}.\label{eq:HQETBox}
\end{align}
We are concerned with addressing how the non-analytic pieces of the integrals in Eqs.~\eqref{eq:QuadBox} and \eqref{eq:HQETBox} are related.

We see from Eq.~\eqref{eq:HQETBox} that the HQET integral is, up to a factor of $1/4m_1m_2$, the leading term of the integral in Eq.~\eqref{eq:QuadBox} when it has been expanded in $\hbar$ or $1/m$ --- the equivalence of the two expansions is once again manifest. However, when including subleading terms in the expansion of Eq.~\eqref{eq:QuadBox}, additional factors of $(l+q)^2$ appear in the numerator, cancelling one of the massless propagators. We conclude that all non-analytic contributions to Eq.~\eqref{eq:QuadBox} must be produced by the leading term of its expansion in $\hbar$ ($1/m$). 
The same argument holds in the cases of triangle and crossed-box integrals, so the non-analytic pieces of integrals with quadratic massive propagators are reproduced (up to a factor of $2m$ for each propagator of mass $m$) by the HQET integrals. Another way of seeing why this should be the case is to invoke generalized unitarity.
Upon cutting two massless propagators $l^2$ and $(l+q)^2$, there is no 
distinction between $I_{\text{quad}}$ and $I_{\text{HQET}}$. Consequently, the one-loop integrals needed to perform the calculations in this paper 
are those in Ref.~\cite{Holstein:2008sw} with $p^{\mu}\rightarrow m v^{\mu}+k^{\mu}$ and multiplied by $2m$ for each massive propagator of mass $m$.\footnote{The integrals in Ref.~\cite{Holstein:2008sw} contain only IR and UV finite terms. It was shown in Ref.~\cite{Bern:2019crd} that the interference of such terms does not contribute to the classical potential, so we have omitted them from our calculations.}

\subsection{Pinch singularity}

HQET box integrals suffer from the so-called pinch singularity, which causes it to be ill-defined and means that HQET cannot be used to describe a bound state of two heavy particles beyond tree level. The cause of this issue is that, in such a scenario, the two heavy particles would have the same velocity, $v_1^\mu = v_2^\mu = v^\mu$. The HQET box integral in Eq.~\eqref{eq:QuadBox} then becomes
\begin{align}
    I_{\text{HQET}} = -\int \frac{d^4 l}{(2\pi)^4} \frac{1}{l^2(l+q^2)\left[v\cdot (l+q) - 
	i \epsilon\right]\left[v\cdot (l+q) + i\epsilon\right]}.\label{eq:PinchSingularity}
\end{align}
Any contour one tries to use to evaluate this integral is then "pinched" in the $\epsilon\rightarrow0$ limit by the singularities above and below the real axis at $v\cdot(l+q)=0$ \cite{Manohar:2000dt}.

For bound systems, the resolution is to reorganize the power counting expansion in terms of $v/c$ instead of $q/m$. The resulting effective theory is non-relativistic QCD (NRQCD), which restores the quadratic pieces of the propagators. In the case at hand, however, we are considering the scattering of two unbound heavy particles, the crucial difference being that the velocities of the heavy particles are in general distinct, ($v_1^\mu \neq v_2^\mu$). Thus, the HQET integral remains well defined.\footnote{We thank Aneesh Manohar for discussions on this point.} Note that the limit where the HQET box integral becomes ill-defined ($v_1^\mu \rightarrow v_2^\mu$) is precisely the limit in which the box integral with quadratic massive propagators obtains the singularity which is removed by the Born iteration \cite{Holstein:2008sw}.

% BIBLIOGRAPHY
% use BIBTEX if you want
%\bibliographystyle{JHEP}
%\bibliography{yourBIBfiles}


\bibliographystyle{JHEP}
\bibliography{hbet}

\providecommand{\href}[2]{#2}\begingroup\raggedright\begin{thebibliography}{10}

\bibitem{LIGOGW}
{B. P. Abbott et al., LIGO Scientific Collaboration and Virgo Collaboration},
  \emph{Observation of gravitational waves from a binary black hole merger},
  \href{https://doi.org/10.1103/PhysRevLett.116.061102}{\emph{Phys. Rev. Lett.}
  {\bfseries 116} (2016) 061102}.

\bibitem{Buonanno:1998gg}
A.~Buonanno and T.~Damour, \emph{{Effective one-body approach to general
  relativistic two-body dynamics}},
  \href{https://doi.org/10.1103/PhysRevD.59.084006}{\emph{Phys. Rev.}
  {\bfseries D59} (1999) 084006}
  [\href{https://arxiv.org/abs/gr-qc/9811091}{{\ttfamily gr-qc/9811091}}].

\bibitem{Damour:2016gwp}
T.~Damour, \emph{{Gravitational scattering, post-Minkowskian approximation and
  Effective One-Body theory}},
  \href{https://doi.org/10.1103/PhysRevD.94.104015}{\emph{Phys. Rev.}
  {\bfseries D94} (2016) 104015}
  [\href{https://arxiv.org/abs/1609.00354}{{\ttfamily 1609.00354}}].

\bibitem{Damour:2017zjx}
T.~Damour, \emph{{High-energy gravitational scattering and the general
  relativistic two-body problem}},
  \href{https://doi.org/10.1103/PhysRevD.97.044038}{\emph{Phys. Rev.}
  {\bfseries D97} (2018) 044038}
  [\href{https://arxiv.org/abs/1710.10599}{{\ttfamily 1710.10599}}].

\bibitem{Baker:2005vv}
J.~G. Baker, J.~Centrella, D.-I. Choi, M.~Koppitz and J.~van Meter,
  \emph{{Gravitational wave extraction from an inspiraling configuration of
  merging black holes}},
  \href{https://doi.org/10.1103/PhysRevLett.96.111102}{\emph{Phys. Rev. Lett.}
  {\bfseries 96} (2006) 111102}
  [\href{https://arxiv.org/abs/gr-qc/0511103}{{\ttfamily gr-qc/0511103}}].

\bibitem{Campanelli:2005dd}
M.~Campanelli, C.~O. Lousto, P.~Marronetti and Y.~Zlochower, \emph{{Accurate
  evolutions of orbiting black-hole binaries without excision}},
  \href{https://doi.org/10.1103/PhysRevLett.96.111101}{\emph{Phys. Rev. Lett.}
  {\bfseries 96} (2006) 111101}
  [\href{https://arxiv.org/abs/gr-qc/0511048}{{\ttfamily gr-qc/0511048}}].

\bibitem{Pretorius:2005gq}
F.~Pretorius, \emph{{Evolution of binary black hole spacetimes}},
  \href{https://doi.org/10.1103/PhysRevLett.95.121101}{\emph{Phys. Rev. Lett.}
  {\bfseries 95} (2005) 121101}
  [\href{https://arxiv.org/abs/gr-qc/0507014}{{\ttfamily gr-qc/0507014}}].

\bibitem{Goldberger:2004jt}
W.~D. Goldberger and I.~Z. Rothstein, \emph{{An Effective field theory of
  gravity for extended objects}},
  \href{https://doi.org/10.1103/PhysRevD.73.104029}{\emph{Phys. Rev.}
  {\bfseries D73} (2006) 104029}
  [\href{https://arxiv.org/abs/hep-th/0409156}{{\ttfamily hep-th/0409156}}].

\bibitem{Porto:2005ac}
R.~A. Porto, \emph{{Post-Newtonian corrections to the motion of spinning bodies
  in NRGR}}, \href{https://doi.org/10.1103/PhysRevD.73.104031}{\emph{Phys.
  Rev.} {\bfseries D73} (2006) 104031}
  [\href{https://arxiv.org/abs/gr-qc/0511061}{{\ttfamily gr-qc/0511061}}].

\bibitem{Porto:2016pyg}
R.~A. Porto, \emph{{The effective field theorist’s approach to gravitational
  dynamics}}, \href{https://doi.org/10.1016/j.physrep.2016.04.003}{\emph{Phys.
  Rept.} {\bfseries 633} (2016) 1}
  [\href{https://arxiv.org/abs/1601.04914}{{\ttfamily 1601.04914}}].

\bibitem{Levi:2018nxp}
M.~Levi, \emph{{Effective Field Theories of Post-Newtonian Gravity: A
  comprehensive review}},  \href{https://arxiv.org/abs/1807.01699}{{\ttfamily
  1807.01699}}.

\bibitem{Donoghue:1993eb}
J.~F. Donoghue, \emph{{Leading quantum correction to the Newtonian potential}},
  \href{https://doi.org/10.1103/PhysRevLett.72.2996}{\emph{Phys. Rev. Lett.}
  {\bfseries 72} (1994) 2996}
  [\href{https://arxiv.org/abs/gr-qc/9310024}{{\ttfamily gr-qc/9310024}}].

\bibitem{Donoghue:1994dn}
J.~F. Donoghue, \emph{{General relativity as an effective field theory: The
  leading quantum corrections}},
  \href{https://doi.org/10.1103/PhysRevD.50.3874}{\emph{Phys. Rev.} {\bfseries
  D50} (1994) 3874} [\href{https://arxiv.org/abs/gr-qc/9405057}{{\ttfamily
  gr-qc/9405057}}].

\bibitem{BjerrumBohr:2002kt}
N.~E.~J. Bjerrum-Bohr, J.~F. Donoghue and B.~R. Holstein, \emph{{Quantum
  gravitational corrections to the nonrelativistic scattering potential of two
  masses}}, \href{https://doi.org/10.1103/PhysRevD.71.069903,
  10.1103/PhysRevD.67.084033}{\emph{Phys. Rev.} {\bfseries D67} (2003) 084033}
  [\href{https://arxiv.org/abs/hep-th/0211072}{{\ttfamily hep-th/0211072}}].

\bibitem{Holstein:2008sx}
B.~R. Holstein and A.~Ross, \emph{{Spin Effects in Long Range Gravitational
  Scattering}},  \href{https://arxiv.org/abs/0802.0716}{{\ttfamily 0802.0716}}.

\bibitem{Bjerrum-Bohr:2013bxa}
N.~E.~J. Bjerrum-Bohr, J.~F. Donoghue and P.~Vanhove, \emph{{On-shell
  Techniques and Universal Results in Quantum Gravity}},
  \href{https://doi.org/10.1007/JHEP02(2014)111}{\emph{JHEP} {\bfseries 02}
  (2014) 111} [\href{https://arxiv.org/abs/1309.0804}{{\ttfamily 1309.0804}}].

\bibitem{Cachazo:2017jef}
F.~Cachazo and A.~Guevara, \emph{{Leading Singularities and Classical
  Gravitational Scattering}},
  \href{https://arxiv.org/abs/1705.10262}{{\ttfamily 1705.10262}}.

\bibitem{Guevara:2017csg}
A.~Guevara, \emph{{Holomorphic Classical Limit for Spin Effects in
  Gravitational and Electromagnetic Scattering}},
  \href{https://doi.org/10.1007/JHEP04(2019)033}{\emph{JHEP} {\bfseries 04}
  (2019) 033} [\href{https://arxiv.org/abs/1706.02314}{{\ttfamily
  1706.02314}}].

\bibitem{GRSA}
N.~E.~J. Bjerrum-Bohr, P.~H. Damgaard, G.~Festuccia, L.~Plant\'{e} and
  P.~Vanhove, \emph{{General Relativity from Scattering Amplitudes}},
  \href{https://doi.org/10.1103/PhysRevLett.121.171601}{\emph{Phys. Rev. Lett.}
  {\bfseries 121} (2018) 171601}
  [\href{https://arxiv.org/abs/1806.04920}{{\ttfamily 1806.04920}}].

\bibitem{Guevara:2018wpp}
A.~Guevara, A.~Ochirov and J.~Vines, \emph{{Scattering of Spinning Black Holes
  from Exponentiated Soft Factors}},
  \href{https://doi.org/10.1007/JHEP09(2019)056}{\emph{JHEP} {\bfseries 09}
  (2019) 056} [\href{https://arxiv.org/abs/1812.06895}{{\ttfamily
  1812.06895}}].

\bibitem{Chung:2018kqs}
M.-Z. Chung, Y.-T. Huang, J.-W. Kim and S.~Lee, \emph{{The simplest massive
  S-matrix: from minimal coupling to Black Holes}},
  \href{https://doi.org/10.1007/JHEP04(2019)156}{\emph{JHEP} {\bfseries 04}
  (2019) 156} [\href{https://arxiv.org/abs/1812.08752}{{\ttfamily
  1812.08752}}].

\bibitem{Zvi3PM}
Z.~Bern, C.~Cheung, R.~Roiban, C.-H. Shen, M.~P. Solon and M.~Zeng,
  \emph{{Scattering Amplitudes and the Conservative Hamiltonian for Binary
  Systems at Third Post-Minkowskian Order}},
  \href{https://arxiv.org/abs/1901.04424}{{\ttfamily 1901.04424}}.

\bibitem{Guevara:2019fsj}
A.~Guevara, A.~Ochirov and J.~Vines, \emph{{Black-hole scattering with general
  spin directions from minimal-coupling amplitudes}},
  \href{https://arxiv.org/abs/1906.10071}{{\ttfamily 1906.10071}}.

\bibitem{Bern:2019crd}
Z.~Bern, C.~Cheung, R.~Roiban, C.-H. Shen, M.~P. Solon and M.~Zeng,
  \emph{{Black Hole Binary Dynamics from the Double Copy and Effective
  Theory}},  \href{https://arxiv.org/abs/1908.01493}{{\ttfamily 1908.01493}}.

\bibitem{Cheung:2018wkq}
C.~Cheung, I.~Z. Rothstein and M.~P. Solon, \emph{{From Scattering Amplitudes
  to Classical Potentials in the Post-Minkowskian Expansion}},
  \href{https://doi.org/10.1103/PhysRevLett.121.251101}{\emph{Phys. Rev. Lett.}
  {\bfseries 121} (2018) 251101}
  [\href{https://arxiv.org/abs/1808.02489}{{\ttfamily 1808.02489}}].

\bibitem{Cristofoli:2019neg}
A.~Cristofoli, N.~E.~J. Bjerrum-Bohr, P.~H. Damgaard and P.~Vanhove, \emph{{On
  Post-Minkowskian Hamiltonians in General Relativity}},
  \href{https://arxiv.org/abs/1906.01579}{{\ttfamily 1906.01579}}.

\bibitem{Kalin:2019rwq}
G.~Kälin and R.~A. Porto, \emph{{From Boundary Data to Bound States}},
  \href{https://arxiv.org/abs/1910.03008}{{\ttfamily 1910.03008}}.

\bibitem{Holstein:2004dn}
B.~R. Holstein and J.~F. Donoghue, \emph{{Classical physics and quantum
  loops}}, \href{https://doi.org/10.1103/PhysRevLett.93.201602}{\emph{Phys.
  Rev. Lett.} {\bfseries 93} (2004) 201602}
  [\href{https://arxiv.org/abs/hep-th/0405239}{{\ttfamily hep-th/0405239}}].

\bibitem{Georgi:1990um}
H.~Georgi, \emph{{An Effective Field Theory for Heavy Quarks at Low-energies}},
  \href{https://doi.org/10.1016/0370-2693(90)91128-X}{\emph{Phys. Lett.}
  {\bfseries B240} (1990) 447}.

\bibitem{Bodwin:1994jh}
G.~T. Bodwin, E.~Braaten and G.~P. Lepage, \emph{{Rigorous QCD analysis of
  inclusive annihilation and production of heavy quarkonium}},
  \href{https://doi.org/10.1103/PhysRevD.55.5853,
  10.1103/PhysRevD.51.1125}{\emph{Phys. Rev.} {\bfseries D51} (1995) 1125}
  [\href{https://arxiv.org/abs/hep-ph/9407339}{{\ttfamily hep-ph/9407339}}].

\bibitem{HQETRev}
M.~Neubert, \emph{{Heavy quark effective theory}}, {\emph{Subnucl. Ser.}
  {\bfseries 34} (1997) 98}
  [\href{https://arxiv.org/abs/hep-ph/9610266}{{\ttfamily hep-ph/9610266}}].

\bibitem{Kosower:2018adc}
D.~A. Kosower, B.~Maybee and D.~O'Connell, \emph{{Amplitudes, Observables, and
  Classical Scattering}},  \href{https://arxiv.org/abs/1811.10950}{{\ttfamily
  1811.10950}}.

\bibitem{Holstein:2008sw}
B.~R. Holstein and A.~Ross, \emph{{Spin Effects in Long Range Electromagnetic
  Scattering}},  \href{https://arxiv.org/abs/0802.0715}{{\ttfamily 0802.0715}}.

\bibitem{Neill:2013wsa}
D.~Neill and I.~Z. Rothstein, \emph{{Classical Space-Times from the S Matrix}},
  \href{https://doi.org/10.1016/j.nuclphysb.2013.09.007}{\emph{Nucl. Phys.}
  {\bfseries B877} (2013) 177}
  [\href{https://arxiv.org/abs/1304.7263}{{\ttfamily 1304.7263}}].

\bibitem{Maybee:2019jus}
B.~Maybee, D.~O'Connell and J.~Vines, \emph{{Observables and amplitudes for
  spinning particles and black holes}},
  \href{https://arxiv.org/abs/1906.09260}{{\ttfamily 1906.09260}}.

\bibitem{Chung:2019duq}
M.-Z. Chung, Y.-t. Huang and J.-W. Kim, \emph{{From quantized spins to rotating
  black holes}},  \href{https://arxiv.org/abs/1908.08463}{{\ttfamily
  1908.08463}}.

\bibitem{ManoharHQET}
A.~V. Manohar, \emph{{The HQET / NRQCD Lagrangian to order alpha / $m^{-3}$}},
  \href{https://doi.org/10.1103/PhysRevD.56.230}{\emph{Phys. Rev.} {\bfseries
  D56} (1997) 230} [\href{https://arxiv.org/abs/hep-ph/9701294}{{\ttfamily
  hep-ph/9701294}}].

\bibitem{Schwartz:2013pla}
M.~D. Schwartz, \emph{{Quantum Field Theory and the Standard Model}}. Cambridge
  University Press, 2014.

\bibitem{Guth:2014hsa}
A.~H. Guth, M.~P. Hertzberg and C.~Prescod-Weinstein, \emph{{Do Dark Matter
  Axions Form a Condensate with Long-Range Correlation?}},
  \href{https://doi.org/10.1103/PhysRevD.92.103513}{\emph{Phys. Rev.}
  {\bfseries D92} (2015) 103513}
  [\href{https://arxiv.org/abs/1412.5930}{{\ttfamily 1412.5930}}].

\bibitem{Braaten:2018lmj}
E.~Braaten, A.~Mohapatra and H.~Zhang, \emph{{Classical Nonrelativistic
  Effective Field Theories for a Real Scalar Field}},
  \href{https://doi.org/10.1103/PhysRevD.98.096012}{\emph{Phys. Rev.}
  {\bfseries D98} (2018) 096012}
  [\href{https://arxiv.org/abs/1806.01898}{{\ttfamily 1806.01898}}].

\bibitem{NRRealScalar}
M.~H. Namjoo, A.~H. Guth and D.~I. Kaiser, \emph{{Relativistic Corrections to
  Nonrelativistic Effective Field Theories}},
  \href{https://arxiv.org/abs/1712.00445}{{\ttfamily 1712.00445}}.

\bibitem{VierGR}
J.~Yepez, \emph{{Einstein's vierbein field theory of curved space}},
  \href{https://arxiv.org/abs/1106.2037}{{\ttfamily 1106.2037}}.

\bibitem{Luke:1992cs}
M.~E. Luke and A.~V. Manohar, \emph{{Reparametrization invariance constraints
  on heavy particle effective field theories}},
  \href{https://doi.org/10.1016/0370-2693(92)91786-9}{\emph{Phys. Lett.}
  {\bfseries B286} (1992) 348}
  [\href{https://arxiv.org/abs/hep-ph/9205228}{{\ttfamily hep-ph/9205228}}].

\bibitem{CHEN1993421}
Y.-Q. Chen, \emph{On the reparameterization invariance in heavy quark effective
  theory},
  \href{https://doi.org/https://doi.org/10.1016/0370-2693(93)91018-I}{\emph{Physics
  Letters B} {\bfseries 317} (1993) 421 }.

\bibitem{Finkemeier:1997re}
M.~Finkemeier, H.~Georgi and M.~McIrvin, \emph{{Reparametrization invariance
  revisited}}, \href{https://doi.org/10.1103/PhysRevD.55.6933}{\emph{Phys.
  Rev.} {\bfseries D55} (1997) 6933}
  [\href{https://arxiv.org/abs/hep-ph/9701243}{{\ttfamily hep-ph/9701243}}].

\bibitem{Bini:2018ywr}
D.~Bini and T.~Damour, \emph{{Gravitational spin-orbit coupling in binary
  systems at the second post-Minkowskian approximation}},
  \href{https://doi.org/10.1103/PhysRevD.98.044036}{\emph{Phys. Rev.}
  {\bfseries D98} (2018) 044036}
  [\href{https://arxiv.org/abs/1805.10809}{{\ttfamily 1805.10809}}].

\bibitem{Arkani-Hamed:2017jhn}
N.~Arkani-Hamed, T.-C. Huang and Y.-t. Huang, \emph{{Scattering Amplitudes For
  All Masses and Spins}},  \href{https://arxiv.org/abs/1709.04891}{{\ttfamily
  1709.04891}}.

\bibitem{Manohar:2000dt}
A.~V. Manohar and M.~B. Wise, \emph{{Heavy quark physics}}, {\emph{Camb.
  Monogr. Part. Phys. Nucl. Phys. Cosmol.} {\bfseries 10} (2000) 1}.

\end{thebibliography}\endgroup

% Please avoid comments such as "For a review'', "For some examples",
% "and references therein" or move them in the text. In general,
% please leave only references in the bibliography and move all
% accessory text in footnotes.

% Also, please have only one work for each \bibitem.

\end{document}